\documentclass[authoryear]{elsarticle}

\usepackage{amsmath,amssymb, amsfonts}
\usepackage{verbatim}
\usepackage{graphicx}
\usepackage{graphics}
\usepackage{geometry}
\usepackage{booktabs}
\usepackage{latexsym}
\usepackage{mathrsfs}
\usepackage{graphicx}
\usepackage{amsmath}
\usepackage{amssymb}
\usepackage{amsfonts}
\usepackage{verbatim}
\usepackage{setspace}

\newtheorem{proposition}{Proposition}[section]

\newtheorem{corollary}{Corollary}[section]

\numberwithin{equation}{section}

\journal{Arxiv}

\begin{document}
\begin{frontmatter}

\markboth{}{}

\title{A Noisy Principal Component Analysis for Forward Rate Curves}

\tnotetext[ML]{We would like to thank Josef Teichmann for stimulating discussions on the topic of this paper. The second author was supported by CNPq grant 308742.}

\cortext[cont]{ Tel.: +55-16-33290867}
\author[eu]{M\'arcio Poletti  Laurini\corref{cont}}%\corref{cont}
\ead[eu]{mplaurini@gmail.com}
\author[ML]{Alberto Ohashi}
\ead[ML]{ohashi@mat.ufpb.br}
\address[eu]{FEA-RP USP - 14040-905, Ribeirão Preto, SP, Brasil}
\address[ML]{Mathematics Dept., Universidade Federal da Para\'iba, 13560-970, Jo\~ao Pessoa, PB, Brasil}

%\thanks{We would like to thank Josef Teichmann for stimulating discussions on the topic of this paper. The second author was supported by CNPq grant 308742.}
%
%\address{Av. dos Bandeirantes 3900,  14040-905, Ribeir\~ao Preto, SP, Brazil} \email{mplaurini@gmail.com}
%
%
%\author{Alberto Ohashi}
%
%\address{Departamento de Matem\'atica, Universidade Federal da Para\'iba, 13560-970, Jo\~ao Pessoa - Para\'iba, Brazil}\email{alberto.ohashi@pq.cnpq.br; ohashi@mat.ufpb.br}
%
%\thanks{We would like to thank Josef Teichmann for stimulating discussions on the topic of this paper. The second author was supported by CNPq grant 308742.}
%\date{\today}

%\keywords{Principal component analysis, term-structure of interest rates.}

%\subjclass{Primary: Secondary:}

\begin{abstract}
Principal Component Analysis (PCA) is the most common nonparametric method for estimating the volatility structure of Gaussian interest rate models. One major difficulty in the estimation of these models is the fact that forward rate curves are not directly observable from the market so that non-trivial observational errors arise in any statistical analysis. In this work, we point out that the classical PCA analysis is \textit{not} suitable for estimating factors of forward rate curves due to the presence of measurement errors induced by market microstructure effects and numerical interpolation. Our analysis indicates that the PCA based on the long-run covariance matrix is capable to extract the true covariance structure of the forward rate curves in the presence of observational errors. Moreover, it provides a significant reduction in the pricing errors due to noisy data typically founded in forward rate curves.
\end{abstract}


\end{frontmatter}

\section{Introduction\label{intro}}

The term-structure of interest rates is a high-dimensional object which has been the subject of much research in the finance literature. It is the natural starting point for pricing fixed-income securities and other financial assets. In particular, the identification of factors capable to explain its movements plays a crucial role in modeling complex interest rate derivative products. Since the seminal works of \cite{steeley}, \cite{stam} and \cite{litterman}, it is well-known that most of the covariance yield curve structure can be summarized by just a few unobservable factors. This stylized fact is fundamentally based on the Principal Component Analysis (henceforth abbreviated by PCA) based on sample covariance matrices. In this case, a small number of eigenvectors summarizes the whole second moment structure of the yield curves.

%Specifically, three factors, which are often called "level", "slope", and "curvature" explain very well the covariance structure of yield curves.

The interest rate markets can be summarized by two fundamental high dimensional objects: the yield $x\mapsto y_t(x)$ and forward rate curves $x\mapsto r_t(x);~t\ge 0$ which are connected by the following linear relation

\begin{equation}\label{1to1}
y_t(x) = \frac{1}{x}\int_0^xr_t(z)dz; 0\le t < \infty,~x\ge 0.
\end{equation}
See e.g.~\cite{key-14} for more details. In particular, the underlying covariance structure of yield and forward rate curves play a major role in the statistical analysis of the term-structure of interest rate. See~e.g.~\cite{rebonatto}, \cite{schmidt} and other references therein. For instance, forward rate curves play a central role in pricing and hedging interest rate derivatives by means of the classical methodology proposed by \cite{heath}. Their contribution can be summarized by the representation of the forward rate curve dynamics in terms of a stochastic partial differential equation

\begin{equation}\label{hjmeq}
dr_t(x) = \Big(  \frac{\partial r_t(x)}{\partial x} + \alpha_{HJM}(t,r_t(x))\Big)dt + \sum_{j=1}^d\sigma_j(t,r_t(x))dB^j_t;~r_0(x) = \xi(x),~0\le t<\infty,~x\ge 0
\end{equation}
where $\alpha_{HJM}$ is the so-called (HJM) drift condition which is fully determined by the volatility structure $\sigma=(\sigma^1, \ldots, \sigma^d)$, $(B^1,\ldots, B^d)$ is a $d$-dimensional Brownian motion and $x\mapsto \xi(x)$ is a given initial forward rate curve. Then, the initial forward rate curve $\xi$ and the volatility structure $\sigma$ fully determine the no-arbitrage dynamics of the model~(\ref{hjmeq}). See~e.g~\cite{key-14} for further details.

Plenty of spot interest rate data (and hence yield curve data) are available in fixed income markets. However, due to the absence of explicit forward rate markets, implied forward rate curves have to be estimated from interest rates based on other financial instruments. This already presents a major difficulty in implementing derivative pricing models based on the classical Heath-Jarrow-Morton methodology.

%In this case, the estimation of forward rates and volatility parameters is done with a number of different methods, some rather complex in order to achieve sufficient precision given the high-dimensional setup. Therefore, a concrete description of the factor structure of forward rate curves is equally important and challenging (see e.g. Alexander~\cite{alexander2}, Rebonatto~\cite{rebonatto} and other references therein).

The most common non-parametric procedure for estimating the covariance structure of forward rate curves is the PCA methodology. Basically, three common strategies are very popular in the PCA estimation of the forward rate curves: \textbf{(A)} One postulates the existence of a finite-dimensional parameterized family of smooth curves $\mathcal{G} = \{G(z; x);~z\in \mathcal{Z}\subset \mathbb{R}^N,x\ge 0\}$ and a $\mathcal{Z}$-valued state process $Y$ such that

\begin{equation}\label{family}
y_t(x) = G(Y_t;x)~\text{for}~x\ge 0,~0\le t <\infty.
\end{equation}
By interpolating the available yield data based on~$\mathcal{G}$, then one extracts the associated forward rate curve by means of any numerical scheme to recover $r_t(x) = y_t(x) - x\frac{\partial y_t(x)}{\partial x}$. The PCA is then applied on this estimated forward rate curves, as discussed in e.g. \cite{key-13} and \cite{lord}. \textbf{(B)} Instead of~(\ref{family}), one shall use a non-parametric polynomial splines method to interpolate the yield data and do the same step of~\textbf{(A)}. See e.g.  \cite{vasicek},  \cite{barzanti} and \cite{chiu} for further details.
Alternatively, one can use proxies to construct the forward rate curve jointly with a given interpolating family of smooth curves $\mathcal{G}$. See e.g. \cite{bhar}), \cite{alexander1} and \cite{gauthier} for further details.

%The covariance structure of forward rate curves plays a major role in Macroeconomic policies, risk management, pricing and hedging interest rate derivatives.

For a given initial forward rate curve, the fundamental object which encodes the whole dynamics of~(\ref{hjmeq}) is volatility. In particular, due to closed form expressions for derivative prices and hedging, it is common (see e.g.  \cite{rutkowski},  \cite{key-13} and \cite{falini}) to assume that the volatility structure is deterministic. In this case, the stochastic dynamics of forward rates is given by a Gaussian HJM model:
\begin{equation}\label{ghjm}
dr_t(x) = \Big(  \frac{\partial r_t(x)}{\partial x} + \alpha_{HJM}(x)\Big)dt + \sum_{j=1}^d\sigma_j(x)dB^j_t;~r_0(x) = \xi(x),~0\le t<\infty,~x\ge 0.
\end{equation}

The most common alternative to estimate the underlying volatility structure is to use PCA methodology (see e.g. \cite{key-13} \cite{key-14}, \cite{schmidt} and other references therein) based on the static covariance matrix of a given sample $(r_t(x_1),\ldots, r_t(x_n))$. The PCA methodology provides the following estimator for the volatility structure

\begin{equation}\label{volpca}
\hat{\sigma}^i = \hat{\varphi}_i \sqrt{\hat{\lambda}_i};~i=1,\ldots, \hat{m},
\end{equation}
where $\hat{m}$ is the estimated number of principal components of the forward rate curves and the estimated eigenvalues and eigenvectors of the associated static covariance matrix are given by $\hat{\lambda}_i$ and $\hat{\varphi}_i$, respectively. One fundamental assumption behind~(\ref{volpca}) and, more generally, on the use of PCA in forward rate curves is the following one:

%Generically, one has to assume weak stationarity of the data, and more importantly, there is no effective observational error in the data.

\

\noindent \textbf{Assumption (I)} There is no observational errors in forward rate curves.

\

%Then it is not hard to see that observational errors may have a non-trivial impact on pricing and hedging interest rate derivatives.

At this stage, a natural question is the validity of assumption \textbf{(I)} in the term-structure of interest rates. In fact, we shall compare the existing literature of principal components between yield and forward rate curves to see some evidence of violation of~assumption~\textbf{(I)}. In one hand, the linearity of the relation~(\ref{1to1}) strongly suggests that dimension of the forward rate and yield curves must be identical (See Proposition~\ref{numberres}). On the other hand, apparently, distinct results in the literature have been reported on the spectral structure of the forward rate and yield curves. \cite{akahori} and \cite{liu} report a remarkable difference in the estimated number of factors between forward rate and yield curves and they suggest that a possible explanation for this would be the violation of the random walk hypothesis. The same type of behavior was reported by \cite{lekkos} who argues that averaging the forward rates over time to maturities would induce a strong dependence on the yield data. He argues that PCA method artificially estimates a small number of principal components for yield curves. \cite{alexander1} study statistical properties of the UK Libor rates. They show that the strategy~\textbf{(A)} induces equivalent loading factor structures between yield and implied forward rate curves. \cite{lord} report a visible difference in the PCA of forward and yield curves by using estimated Svensson curves for the Euro market.

%They conclude that the choice of the yield curve fitting technique~(see~(\ref{family})) affects the forward rate covariance structure much more than the choice of sample size.

Essentially, the existing literature restricts the discussion into two lines: (i) More factors are needed to account the correlation in forward rates curves. An averaging effect would be the reason for an artificial dependence on the yield curves. (ii) One way to remedy this pattern is to first smooth the yield data by means of a parametric or non-parametric form and then to calculate the implied forward rates. One should notice that one important assumption behind (i) and (ii) is \textbf{(I)}. More importantly, the current literature only suggests \textit{ad hoc} methods based on \textbf{(A-B)}.

In this article, we take a rather different strategy. Throughout this article, we assume that the observed curve time series, which we denote by $X_1(\cdot), \ldots, X_n(\cdot)$, they are subject to errors in the sense that

\begin{equation}\label{errorOBS}
X_t(u) = r_t(u) + \varepsilon_t(u);~u\ge 0, t=0,1, \ldots~.
\end{equation}
The possible existence of the noise term $\varepsilon$ in (\ref{errorOBS}) reflects the classical interpolation procedures~\textbf{(A-B)} when extracting forward rate curves from yield data. It can also be induced by observational errors due to market microstructure effects. The existence of an underlying bid and ask bond price structure contaminates the yield and forward rate curves~(see e.g~\cite{Mizrach} and \cite{JFQ}). These noisy discrete data are smoothed to provide ``observed" curves $x\mapsto X_t(x)$ where both $r_t(\cdot)$ and $\varepsilon_t(\cdot)$ are unobservable. We investigate in detail the existence and the impact of observational errors $\{\varepsilon_t(x); t\in \mathbb{N},x\in \mathbb{R}_+\}$ in~(\ref{errorOBS}) in the classical PCA methodology.

We show that market microstructure effects and common interpolation procedures~\textbf{(A-B)} induce noisy forward rate curves which cause severe bias in the PCA methodology. The starting point of our analysis is the fact that the ranks of the covariance operators of the forward rate and yield curves are identical (see Proposition~\ref{numberres}). In addition, we show that PCA based on the so-called \textit{long-run covariance matrix} (henceforth abbreviated by LRCM) significantly improves the estimation of the covariance structure of forward rate curves. The impact of noisy data in pricing interest rate derivatives is also discussed.

This article is organized as follows. In Section 2, we report an elementary result about the equivalence of ranks between the covariance operators of the forward rate and yield curves. In Section 3, we describe some alternatives of estimating covariance structures in the presence of observational errors, the so-called LRCM estimators. Section 4 presents a detailed simulation analysis reporting the performance of the PCA based on LRCM estimators, as well as, the role played by measurement errors in the PCA methodology applied to the term-structure of interest rate. In order to compare the simulation results with a real data set, Section 5 provides an empirical analysis on the number of principal components for US and UK term-structure on interest-rates. In Section 6, we analyse the impact of neglecting observational errors in pricing interest rate derivatives in light of the PCA methodology. Section 7 presents the final remarks.

\section{Rank equivalence in covariance operators for forward rate and yield curves}
In this section, we give a simple remark showing the number of principal components in forward rate curves should be exactly the same of the yield curves under some mild conditions. In spite of its simplicity, it is the starting point to investigate the violation of assumption~\textbf{(I)} and it also gives a comparison criteria for our statistical analysis.

Let $\{P(t,T); (t,T)\in \Delta^2 \}$ be the term-structure of bond prices where $\Delta^2:=\{(t,T); 0\le t\le T < \infty \}$. Let

\begin{equation}\label{yld}
s(t,T):= \frac{-log~P(t,T)}{T-t}; (t,T)\in \Delta^2,
\end{equation}
be the spot-interest rate prevailing at time $t$ for maturity $T$ and let $y_t(x):= s(t,t+x)$ be the correspondent yield curve at time indexed by the time to maturity $x=T-t$.

\noindent The forward rate prevailing at time $t$ for maturity $T$ is

$$
f(t,T):= -\frac{\partial~log P(t,T)}{\partial T}; (t,T)\in \Delta^2,
$$
and the forward rate curve is $r_t(x):= f(t,t+x)$ for $x=T-t$. For simplicity of exposition, we analyze the PCA methodology based on a space $E$ of curves so that forward rate and yield data are interpreted as sample curves over time.

In the sequel, we consider a discrete-time setup $t\in \mathbb{N}:=\{0,1, 2,  \ldots\}$ and we assume that $ y_t$ and $r_t$ are discrete-time weakly stationary $E$-valued process. That is, there exist functions $(\mu_y(\cdot), \mu_r(\cdot), Q_r(\cdot,\cdot), Q_y(\cdot,\cdot))$ such that the following identities hold for every $t$

$$\mu_r(u) = \mathbb{E}r_t(u), \quad \mu_y(u) = \mathbb{E}y_t(u),$$

\begin{equation}\label{kernelcov}
\quad Q_y(u,v)=Cov(y_t(u),y_t(v)), \quad Q_r(u,v)=Cov(r_t(u), r_t(v));~u,v\in K.
\end{equation}
Otherwise, we assume that the first difference process satisfies such properties. The set $E$ is a separable Hilbert space of functions from a bounded set $K:=[0,x^\star)\subset \mathbb{R}_+$ to $\mathbb{R}$ such that $J,T:E\rightarrow E$ defined by

\begin{equation}\label{boundedness}
f\mapsto J_x(f):=\frac{1}{x}\int_0^xf(y)dy; \quad f\mapsto T_x(f):= x\frac{d}{d x}f(x),
\end{equation}
are bounded linear operators. In the sequel, we denote $E$ equipped with an inner product $\langle \cdot,\cdot \rangle^{1/2}=\|\cdot\|$. Assuming that the discrete-time processes $r$ and $y$ are square-integrable, the covariance operators induced by the kernels in~(\ref{kernelcov}) admit spectral decompositions over $E$ as follows

$$Q_y(\cdot) = \sum_{k=1}^\infty \lambda_k(y)\langle \cdot, \varphi_k(y)\rangle \varphi_k(y),\quad Q_r(\cdot) = \sum_{k=1}^\infty \lambda_k(r)\langle \cdot, \varphi_k(r)\rangle \varphi_k(r),$$
where $(\varphi_k(r))_{k=1}^\infty$ and $(\varphi_k(y))_{k=1}^\infty$ are orthonormal bases for $E$ with eigenvalues $(\lambda_k(r))_{k=1}^\infty$ and $(\lambda_k(y))_{k=1}^\infty$, respectively. The number of principal components of the forward and yield curves are the number of non-zero eigenvalues of $Q_r$ and $Q_y$, respectively. We assume that $\lambda_1(r) > \ldots > \lambda_p(r) > \lambda_{p+i}=0$; $\lambda_1(y)>\ldots>\lambda_q(y)> \lambda_{q+i}=0$ for every $i\ge 1$,  $max~(p,q)< \infty$ so that
$$\mathbb{E}\|r_t-\mu_r\|^2=\sum_{i=1}^p \lambda_i(r),\quad \mathbb{E}\| y_t-\mu_y\|^2=\sum_{i=1}^q \lambda_i(y);~ t\in \mathbb{N}.$$

\noindent The explained variance associated with the $k$-th principal component for $r$ and $y$ can be expressed, respectively, by
$$
\frac{\lambda_k(r)}{\sum_{i=1}^p\lambda_i(r)},\quad \frac{\lambda_k(y)}{\sum_{i=1}^q\lambda_i(y)}.
$$
In the sequel, we denote $\Lambda:= I_d - T$ where $I_d$ is the identity operator and $T$ is given by~(\ref{boundedness}). For a given bounded linear operator $G$, the correspondent self-adjoint operator will be denoted by $G^*$. The following simple remark shows how the covariance operators of the forward rate and yield curves are related to each other.

\begin{proposition}\label{numberres}
Let $y$ and $r$ be square-integrable weakly-stationary $E$-valued discrete-time processes. Assume that $Q_r$ and $Q_y$ are finite-rank operators and $E$ is a Hilbert space realizing~(\ref{boundedness}). Then $Q_r = \Lambda Q_y \Lambda^*$ and $Q_y=JQ_r J^*$. In particular, $Rank~Q_r = Rank~Q_y$.
\end{proposition}

\noindent \textbf{Proof:} We recall that $Q_r$ and $Q_y$ are the unique self-adjoint, non-negative and bounded operators such that

\begin{equation}\label{e5}
\langle Q_r f, g\rangle= \mathbb{E}\langle r_t -\mu_r, f\rangle\langle r_t-\mu_r, g\rangle
\end{equation}

\begin{equation}\label{e6}
\langle Q_y f, g\rangle= \mathbb{E}\langle y_t -\mu_y, f\rangle\langle y_t-\mu_y, g\rangle;~f,g\in  E, t\in \mathbb{N}.
\end{equation}
Moreover, we shall write $y_t(x) = J_x(r_t)$ and $r_t(x) = \Lambda_x ( y_t)$. This fact, together with relations~(\ref{e5}) and~(\ref{e6}) and the continuity assumptions on $\Lambda$ and $J$ allow us to conclude that $Q_r = \Lambda Q_y \Lambda^*$ and $Q_y=JQ_r J^*$. Now let us define the following finite-rank, non-negative and self-adjoint operators

$$\bar{Q}_r:= Q_y\Lambda^* \Lambda, \quad \bar{Q}_y:= Q_r J^* J.$$
By the very definition, $\bar{Q}_y$ and $Q_y$ share the same non-zero eigenvalues and therefore,

$$Rank~Q_y=Rank~\bar{Q}_y\le Rank~Q_r J^*\le Rank~Q_r.$$
The same argument applies to $\bar{Q}_r$ and $Q_r$ so that $Rank~Q_r \le Rank~Q_y$. This concludes the proof.

\

Despite the simplicity of Proposition~\ref{numberres}, it provides an important information on dimension reducing techniques for forward rate and yield curves based on PCA: The effective number of principal components should be the same for forward rates and yield curves.

%Eventually, the spectral structure induced by $Q_r$ and $Q_y$ may be commonly represented by the well-known level-curvature-slope phenomenon\footnote{For instance, if one has a correlation structure of the form $\rho^{|t-s|}$ ($\rho >0$) then \cite{forzani} show that the loading factors are perturbations of cosine.}. For a discussion on this matter, see \cite{forzani}, \cite{lord} and \cite{reis} and other references therein.

\subsection{Implications to related work}Proposition~\ref{numberres} contradicts the heuristic argument given by \cite{lekkos} who argues that the identity
$y=J(r)$ would smooth the spectral structure of the yield curve covariance operator. It also shed some light on the results reported by many authors who compare the PCA for forward and yield curves. In one hand, \cite{akahori1}, \cite{liu}, \cite{lord} and \cite{lekkos} report substantial differences between the correlation structure of forward and yield curves. On the other hand,  \cite{alexander1} report a very stable estimation by using discretely compounded forward rates and the parametric Svensson family to extract the forward rates. Their empirical results together with Proposition~\ref{numberres} might suggest that in order to extract the spectral properties of forward rates from the observed bond prices, a reasonable choice of a parametric family of smooth curves is a good starting point. However, one should notice that this procedure may potentially introduce an a priory loading factor structure on forward rates. As pointed out by \cite{lord}, nonparametric procedures based on splines and bootstrap techniques might introduce a non-negligible noise on the forward curve.

Proposition~\ref{numberres} implies that a remarkable difference in the number of principal components between forward and yield curves is a strong evidence for the presence of unobserved noise in the data. If this is the case, then unobserved errors might induce a nontrivial bias in the statistical analysis of the forward rate curves. This remark will be the starting point for our analysis in the remainder of this article.

\section{PCA based on LRCM and noisy data}\label{methods}

One way to overcome the presence of observational errors in a PCA methodology is the so-called long-run covariance matrix (LRCM). We recall that the PCA method based on the usual static sample covariance matrix is ​​only valid for independent and weakly stationary processes. In the presence of some sort of temporal dependence caused by serial correlation or a contamination process, the use of the static sample covariance matrix is not the correct one anymore.

One alternative to correct those types of dependence is to use estimators based on the LRCM. In the sequel, we denote the process of interest as a vector $w_t$, and assume weak stationarity and ergodicity of $w_t$. The LRCM of $w$ is defined by

\[
V_{lr}:=\lim_{n\rightarrow\infty}var(\sqrt n \bar w )=\sum_{j=-\infty}^{\infty}\gamma(j),
\]
where $\gamma(j):=E\Big[\big(w_{t}-E[w_{t}]\big)\big(w_{t-j}-E[w_{t-j}]\big)^{\top}\Big]$
is the cross covariance in lag \textit{j}, $\bar{w}$ is the sample mean  and $\top$ denotes the transpose of a matrix. One important and standard case is $\gamma(j)$ at $j=0$. In this case, all the non-contemporaneous cross-variances are equal to zero and we retrieve the usual static covariance matrix $V_s:= \gamma(0)$ whose the usual estimator will be denoted by $\hat{V}_s$.

Although it is possible to estimate independently each covariance term $\gamma(j)$ by the correspondent sample quantities $\hat{\gamma}(j)$, the natural estimator of the long run matrix
$\hat{V}^{n}:=\sum_{j=-(n-1)}^{n-1}\hat{\gamma}(j)$ is not consistent because the number of parameters grows proportional to the sample size. In order to overcome this problem, a general non-parametric class of LRCM consistent estimators is introduced by \citet{andrews}

\begin{equation}\label{maes}
\hat{V}_{lr}^{A}:=\sum_{j=-(n-1)}^{(n-1)}\alpha(j)\hat{\gamma}(j),
\end{equation}
where $\alpha(j)$ is a sequence of weights of the form $\alpha(j)=K(j/b)$, where $K(\cdot)$ is a continuous symmetric kernel function such that $K(0)=1$ and $b$ a suitable bandwidth parameter such that $b\rightarrow\infty$ as $n\rightarrow\infty$.

Some optimal choices for the kernel function and the bandwidth parameter in~(\ref{maes}) are discussed by  \citet{andrews}. The most important one is the bandwith parameter. Precise conditions for consistency of the LRCM estimators of type~(\ref{maes}) based on kernel methods require that the bandwith increases slower than sample size.

However it is well-known that this class of asymptotic estimators does not go well in finite samples, in particular for processes with strong dependence and temporal heterogeneity (see e.g.~\citet{key-4}). As discussed by \citet{key-4}, statistical inference by using consistent LRCM estimators performs badly in small samples and also in the presence of dependence and contamination/measument errors.  To overcome these problems, \citet{key-3,key-2} introduce a class of kernel-type estimators with a bandwidth rule given by a fixed portion of the sample size, known as fixed-b estimators. These LRCM estimators are not consistent due to a fixed bandwidth parameter. However, they explicitly incorporate parameter uncertainty and they present very good finite sample properties in hypothesis testing. See \citet{key-3,key-2} for details. In particular, \citet{key-3} suggest the use of the whole sample as a possible bandwidth rule in the construction of the LRCM estimator~(\ref{maes}). This strategy allows us to use all lags of $X$ in~(\ref{errorOBS}) in the estimation procedure. This will be particularly important for noisy data sets with contaminations induced by market microstructure effects and interpolation procedures \textbf{(A-B)} discussed in the Introduction.

Let $\hat u_t$ be the mean adjusted deviations of the series $w_t$. The \citet{key-3} estimator is defined by

\begin{equation}\label{vk}
\hat{V}_{lr}^{VK}:=T^ {-1}\sum_{i=1}^{T} \sum_{j=1}^{T} \hat u_{i}\left( 1-\frac{|i-j|}{T}\right)\hat u_{j},
\end{equation}
which corresponds to the use of a Bartlett kernel and bandwidth equal to the sample size in the LRCM estimation.

In order to handle dependent error structures and outliers, \citet{key-4} extends the results in~\citet{key-2,key-3}. He shows that the usual LRCM estimators are extremely fragile in the presence of contamination and outliers. He suggests a class of LRCM estimators with bandwidths based on fixed portions of the sample size. The class of estimators proposed by~\citet{key-4} can be constructed in the same spirit of \citet{key-3,key-2} but with one fundamental difference: They are asymptotically robust to contaminations in the autocorrelation structure, in particular to contamination by moving average process. In particular, he obtains a class of estimators $\hat{V}_{lr}^{UA(p)}$ (see pages 1339-1342  in~\citet{key-4}) which trades optimally robustness and efficiency. The $\hat{V}_{lr}^{UA(p)}$ estimator is defined by

\begin{equation}\label{ua}
\hat{V}_{lr}^{UA(p)}:=\frac{\sum_{i=1}^{T} \sum_{j=1}^{T} \hat \xi_{i} \hat \xi_{j}}{p},
\end{equation}
where $\hat \xi_t $ is the residual of the linear regression of $\hat u_t$ against a $p$-dimensional series $\hat \upsilon_t(l)$, $l=1,\dots,p$ where $\hat \upsilon_t(l):=\sqrt{2/T} \cos (l \pi (t-1/2)/T$. The parameter $1\le p< \infty$ controls the bias and efficiency in the LRCM estimation procedure. \citet{key-4} shows that this estimator has good properties w.r.t robustness and efficiency. In addition, they present well-known asymptotic distributions under the null in hypothesis testing.

In this article, we show that the LRCM estimators~(\ref{vk}) and~(\ref{ua}) allow us to filter the dependence structure generated by observational errors and market microstructure effects, typically founded in forward rate markets.

%Moreover, our numerical experiments suggest that the use of those estimators increase the accuracy in pricing interest rate derivatives based on Gaussian HJMmodels.

\section{Simulation study on the number of factors}\label{numericalres}
In order to investigate the impact of observational errors in the classical PCA methodology applied to forward rate curves, we now provide a detailed simulation study as follows. In order to study different types of data contamination in forward rate curves, we assume that both the yield and forward rate curves are subject to errors

\begin{equation}\label{XandZ}
X_t = r_t + \varepsilon_t,\quad Z_t = y_t + \eta_t;~t=0,\ldots, T,
\end{equation}
where $\varepsilon_t$ and $\eta_t$ are the error components to be specified.

In addition to the analysis of the curves~(\ref{XandZ}), it is also important to consider the first-difference of the time series $\tilde{X}_t: = X_t - X_{t-1}$ and $\tilde{Z}_t := Z_t-  Z_{t-1}; t=1,\ldots, T.$ We recall that a direct application of the PCA to the processes~(\ref{XandZ}) implicitly assumes that $X$ and $Z$ are already weakly stationary. If $X$ and (or) $Z$ are assumed to be non-stationary, it is necessary to apply the PCA decomposition at the first-differences $\tilde{X}$ and $\tilde{Z}$, which are possibly weakly stationary processes. There is also an intrinsic reason to study the first-difference of the time series. Due to the HJM representation of the forward rate (see~(\ref{hjmeq})), the PCA methodology must be computed in terms of the increments rather than the first levels. See~\cite{key-13} and~\cite{key-14} for further details.

An obvious consequence of Proposition~\ref{numberres} is the following corollary.

\begin{corollary}\label{corollary1}
Assume that both $r$ and $y$ are independent weakly stationary processes. If $\eta=\varepsilon=0$, then the number of principal components of $X$ and $Z$ must be identical.
\end{corollary}
The number of principal components will be the key parameter of study in order to infer the performance of the usual PCA methodology based on $\hat{V}_s$ against the LRCM estimators defined in~(\ref{maes}),~(\ref{vk}) and~(\ref{ua}). Of course, the same statement of Corollary~\ref{corollary1} holds for the first-difference processes $\tilde{X}$ and $\tilde{Y}$.

In order to investigate the impact of observational errors in the PCA methodology, we proceed the analysis on two different prominent models: Gaussian HJM and Cox-Ingersoll-Ross models. As far as the Gaussian HJM model $r_t$ (see~(\ref{ghjm})) is concerned, the volatility parameter is calibrated based on the classical interest rate curves studied by \citet{key-12}. It consists of zero coupon (Treasury bond) bonds with maturities 3, 6, 9, 12, 15, 18, 21, 24, 30, 36, 48, 60, 72, 90, 108 and 120 months covering 1985-Jan to 2000-Dec. The Gaussian HJM simulation is based on this calibrated volatility parameter where the number of principal components is equal to 3 (three).

The specification of the Cox-Ingersoll-Ross model (henceforth abbreviated by CIR model) is based on \cite{key-15}. In this case, a three-factor CIR model is simulated by means of a short rate process of the form $short_t=\sum_{i=1}^{3}Y_{t}^{i}$ where each latent factor follows the stochastic differential equation

$$dY_{t}^{i}=\kappa_{i}(\theta_{i}-Y_{t}^{i})dt+\sigma_{i}\sqrt{Y_{t}^{i}}dB_{t}^{i};~i=1, 2, 3.$$
The parameters $\kappa_i,\theta_i$ and $\sigma_i$ used in the simulation study are chosen according to \cite{key-15} who estimate them based on weakly data (1980-1988) of the U.S Treasury market\footnote{See \citet{key-15} for the expressions for the affine factors and bond prices in the multifactor CIR model.}. We perform 1,000 replications of these HJM and CIR data generating processes with sample size 1,000.

We analyze the PCA methodology based on the standard sample covariance matrix estimator $\hat{V}_s$ against the LRCM estimators $\hat{V}^A_{lr}$, $\hat{V}^{VK}_{lr} $ and $\hat{V}^{UA(p)}_{lr}$. Those estimators are applied to $(X,Z)$ and $(\tilde{X},\tilde{Z})$. The $\hat{V}_{lr}^{A}$ estimator is specified with the quadratic-spectral kernel function and the optimal bandwidth choice of \citet{andrews}. The $\hat{V}_{lr}^{VK}$ estimator is specified with the Bartlett kernel function and the whole sample as a bandwidth rule. The $\hat{V}_{lr}^{UA(p)}$ estimator is specified with $p=4$ components in the basis $\hat{v}_t$. See Section~\ref{methods} for details.

%The forward rate and yield curves are obtained via~(\ref{1to1}).

Figures \ref{fig:zeroGaussian-HJM}-\ref{fig:splineCox-Ingersoll-Ross} report the mean value of the cumulative R-Squared obtained in the Monte Carlo experiments with the PCA decompositions based on the static covariance matrix and the LRCM estimators. Figure~\ref{fig:zeroGaussian-HJM} reports the PCA methodology applied to $X = r + \varepsilon$ where $\varepsilon = 0$ and the associated yield curve is computed via~(\ref{1to1}). Figure \ref{fig:zeroCox-Ingersoll-Ross} reports the PCA methodology applied to $Z = y + \eta$ where $\eta = 0$ and the associated forward rate curve is computed via $y_t(x) + x\frac{\partial y_t(x)}{\partial x}$. Without the presence of observational errors, we notice that the number of principal components is correctly estimated for $(X,Z)$ and $(\tilde{X},\tilde{Z})$ by using any of the estimators. This result is not surprising due to Proposition~\ref{numberres} and the fact that both $r$ and $y$ are weakly stationary processes. In the presence of observational errors the picture is rather different.

\subsection{PCA with measurement error}
We formulate the analysis based on two types of observational errors: Market microstructure effects (MME) and interpolation error structure (IES) as described by \textbf{(A-B)} in the Introduction. In order to analyze the impact of MME in the PCA methodology, we introduce an additive error structure typically founded in interest rate markets due to transactions costs and liquidity premiums in bond prices (see e.g~\cite{Mizrach} and \cite{JFQ}). This phenomenon is directly observed by the existence of bid and ask prices. In the classical model of market microstructure (e.g. \cite{hasbrouck}), the true price of the asset is within the range between the observed bid and ask prices. In other words, the observed prices may be considered as the true price plus an additive measurement error.

\begin{figure}
\begin{centering}
\includegraphics[width=6cm,height=5cm]{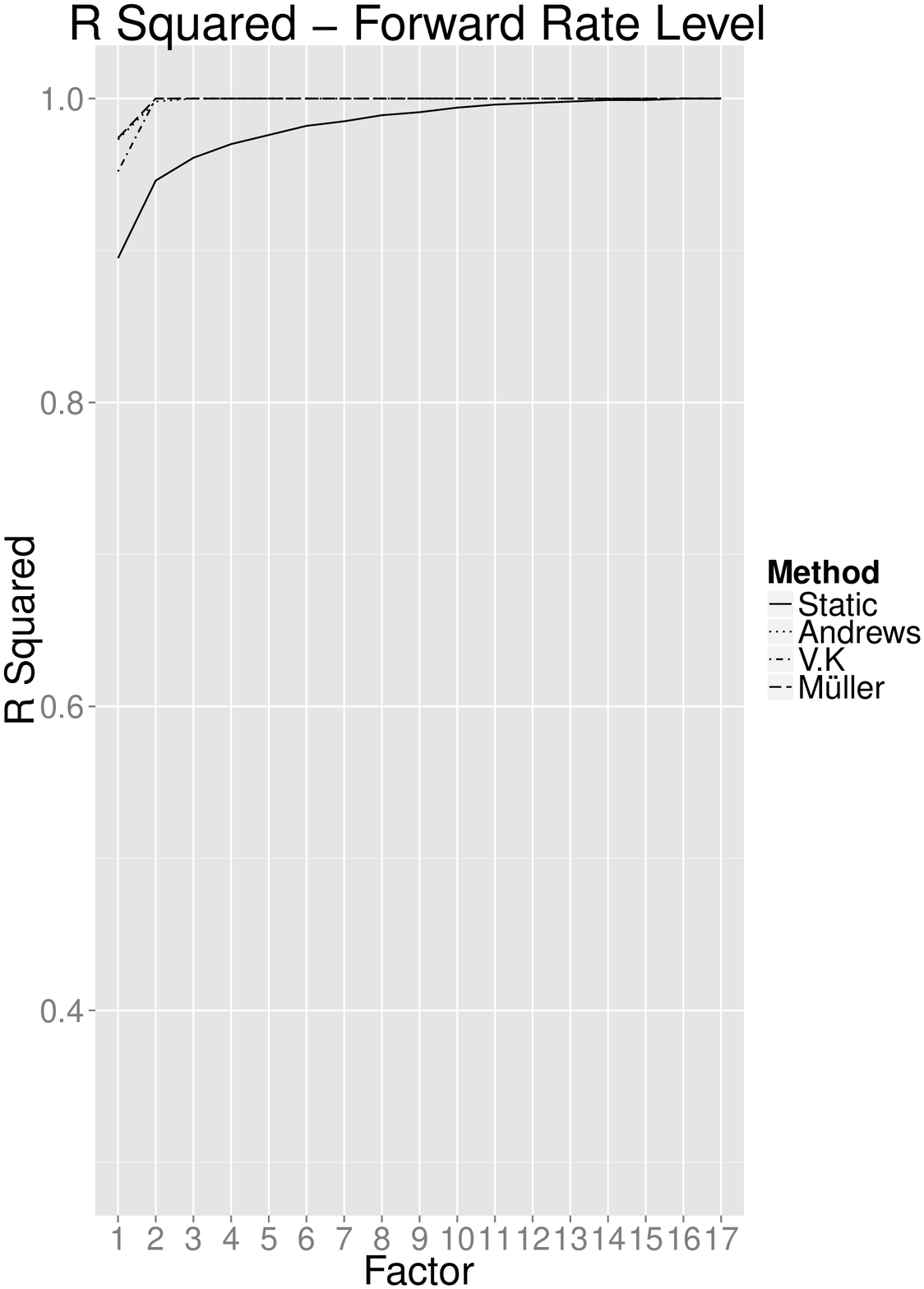}\includegraphics[width=6cm,height=5cm]{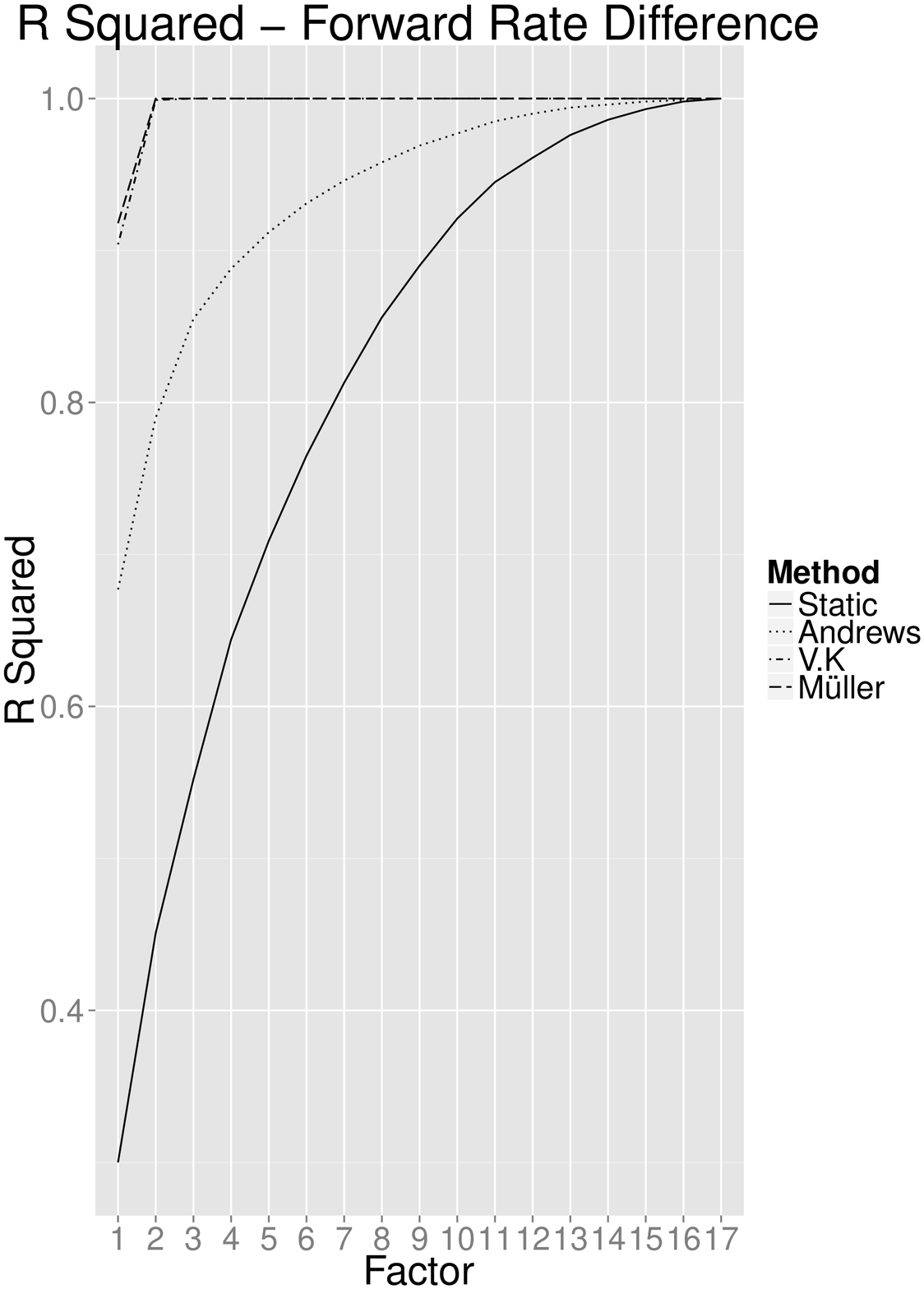}
\par\end{centering}

\begin{centering}
\includegraphics[width=6cm,height=5cm]{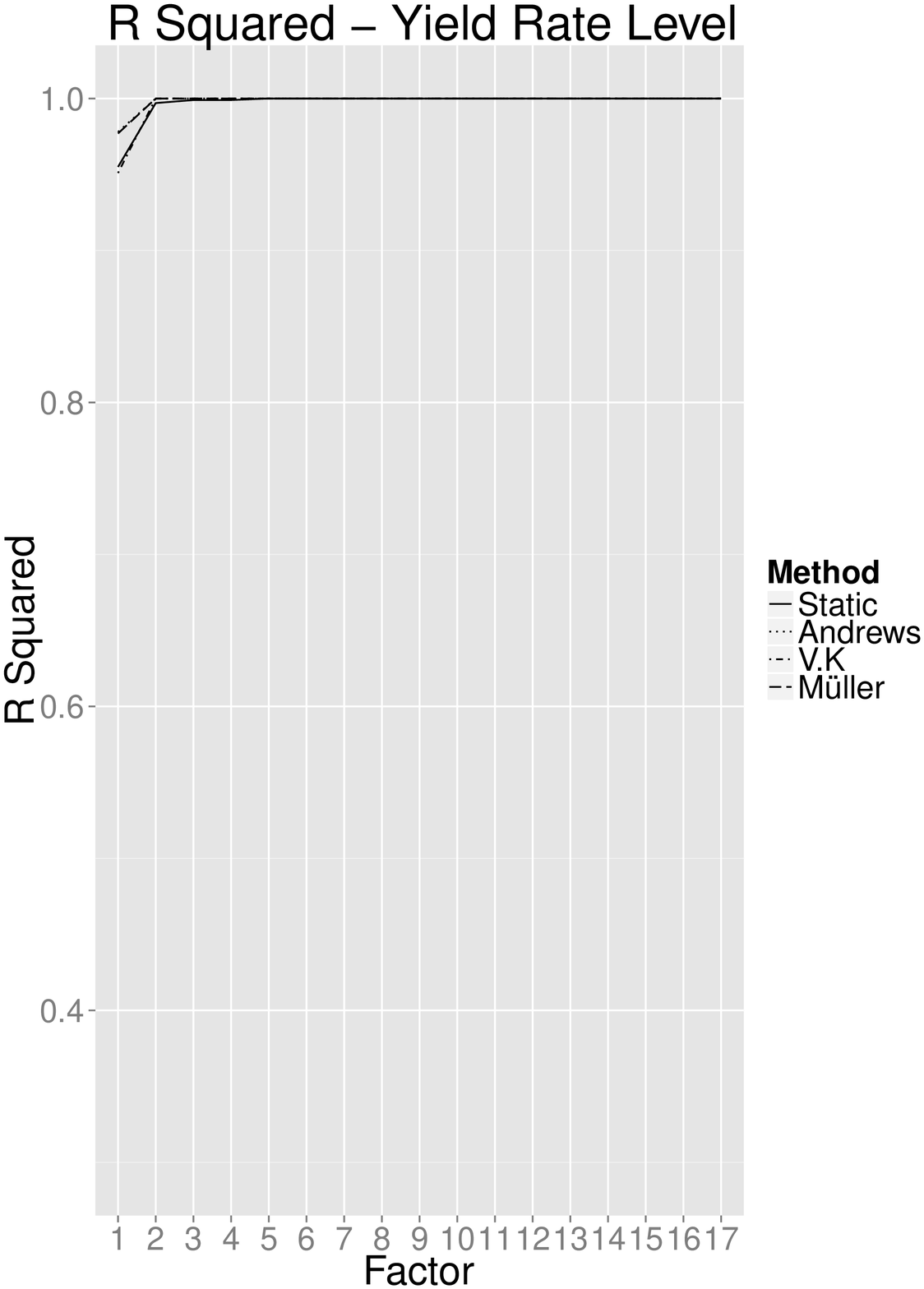}\includegraphics[width=6cm,height=5cm]{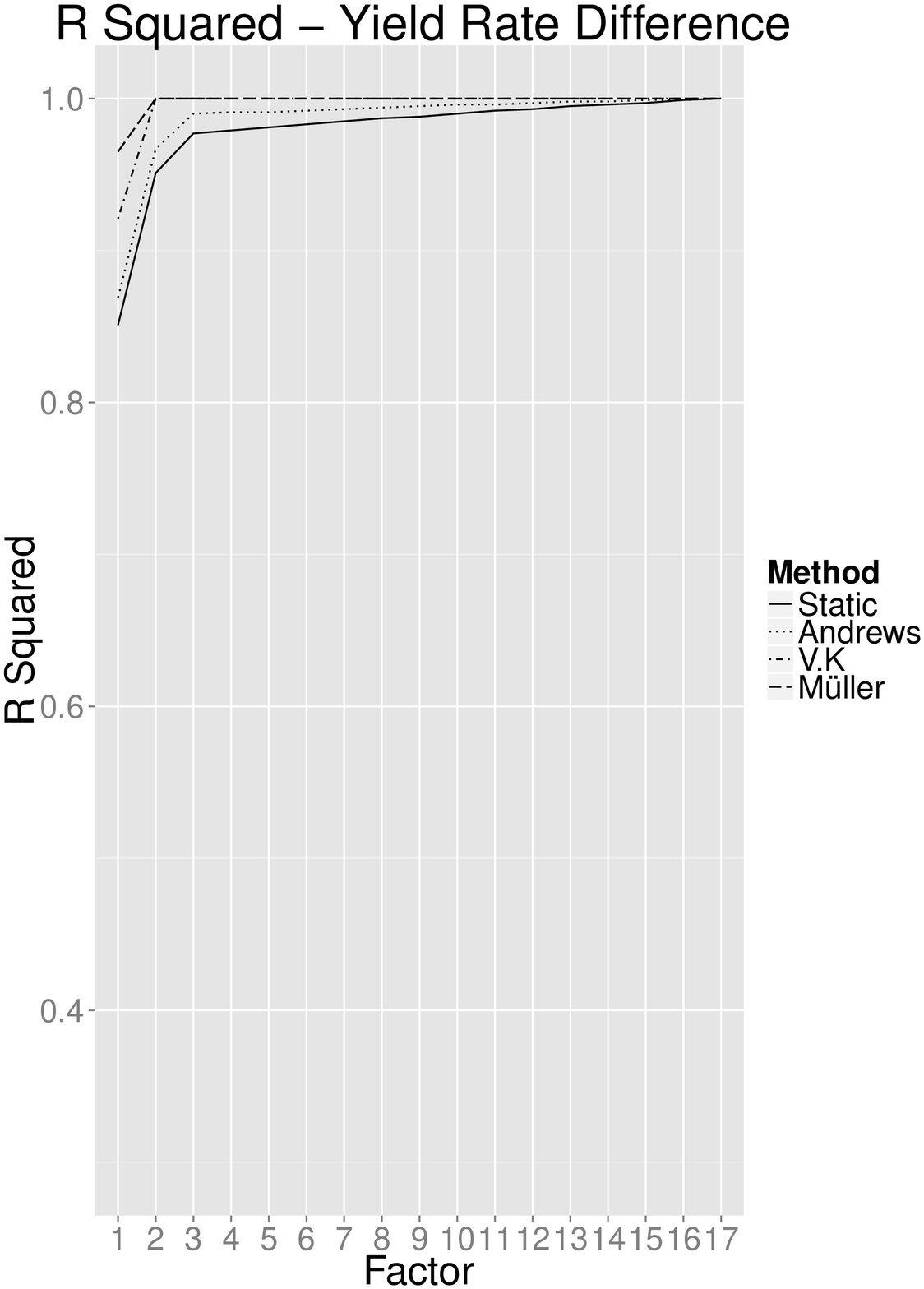}
\par\end{centering}

\begin{centering}
\caption{Cumulative $R^{2}$ obtained from the PCA decomposition for the level and first
difference of the forward and yield term structures. First experiment with  observational errors generated by  market microstructure effects, Gaussian HJM process\label{fig:Gaussian-HJM}. Mean values from
1,000 Monte Carlo simulations.}

\par\end{centering}

\end{figure}

\begin{figure}
\begin{centering}
\includegraphics[width=6cm,height=5cm]{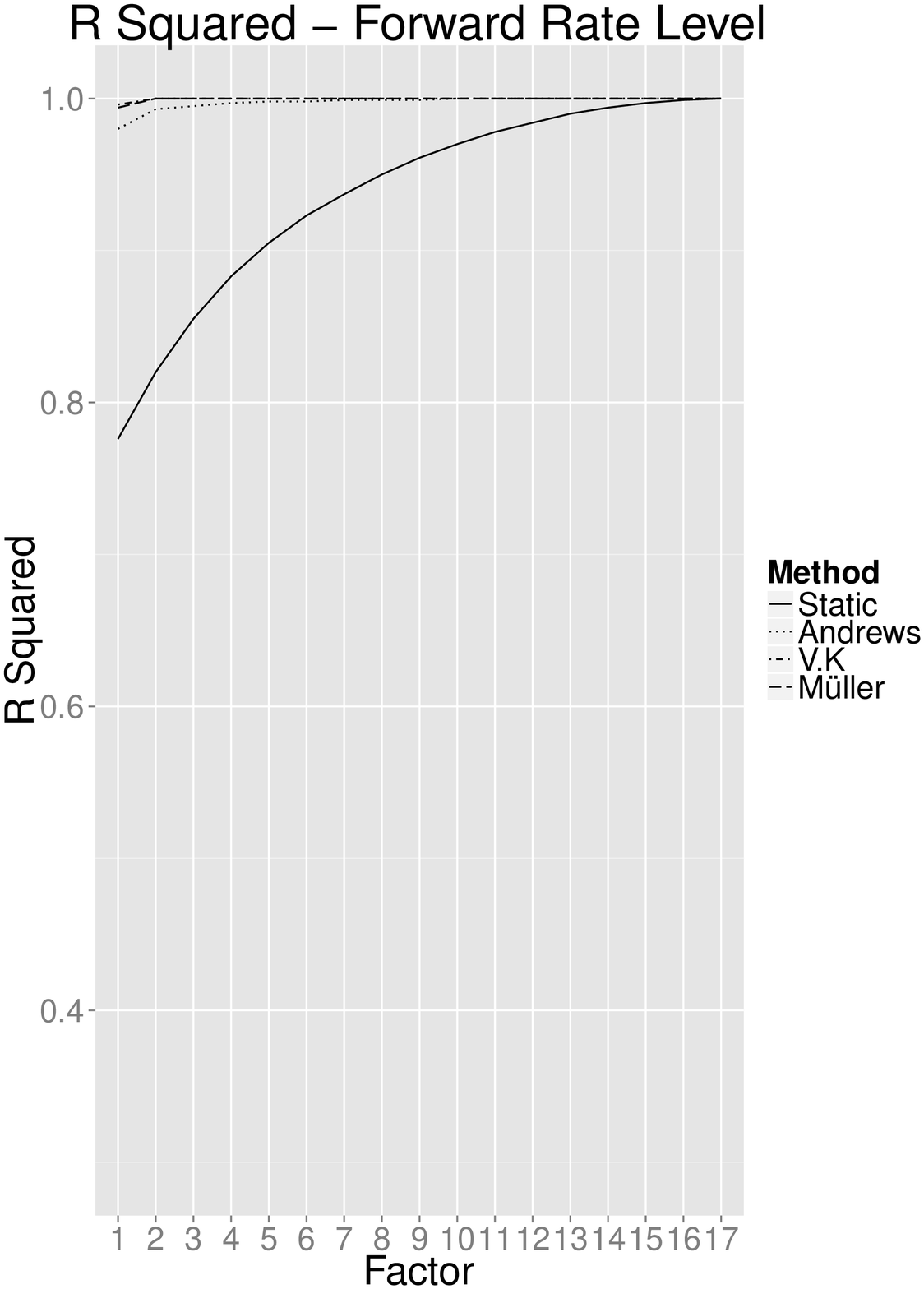}\includegraphics[width=6cm,height=5cm]{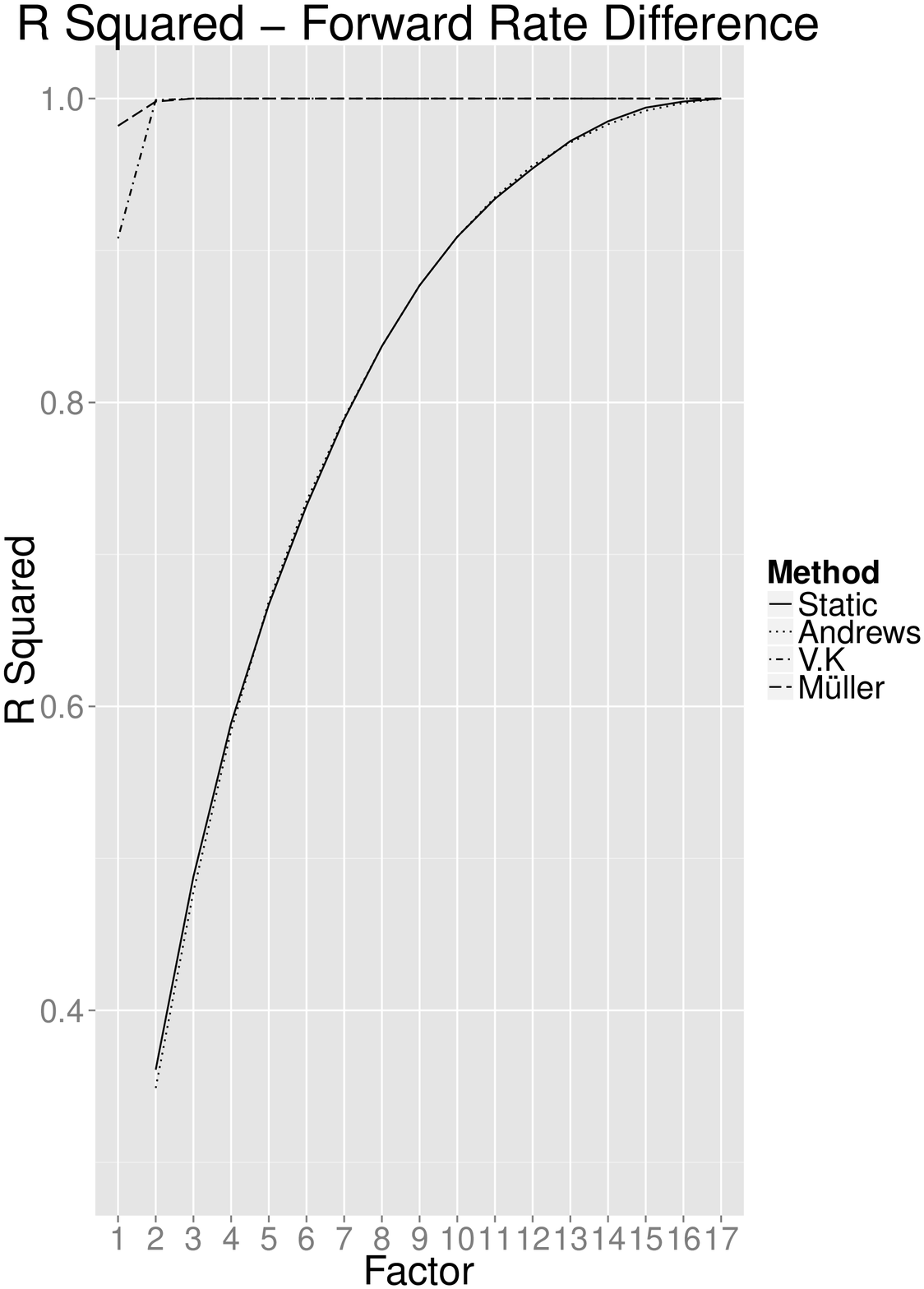}
\par\end{centering}

\begin{centering}
\includegraphics[width=6cm,height=5cm]{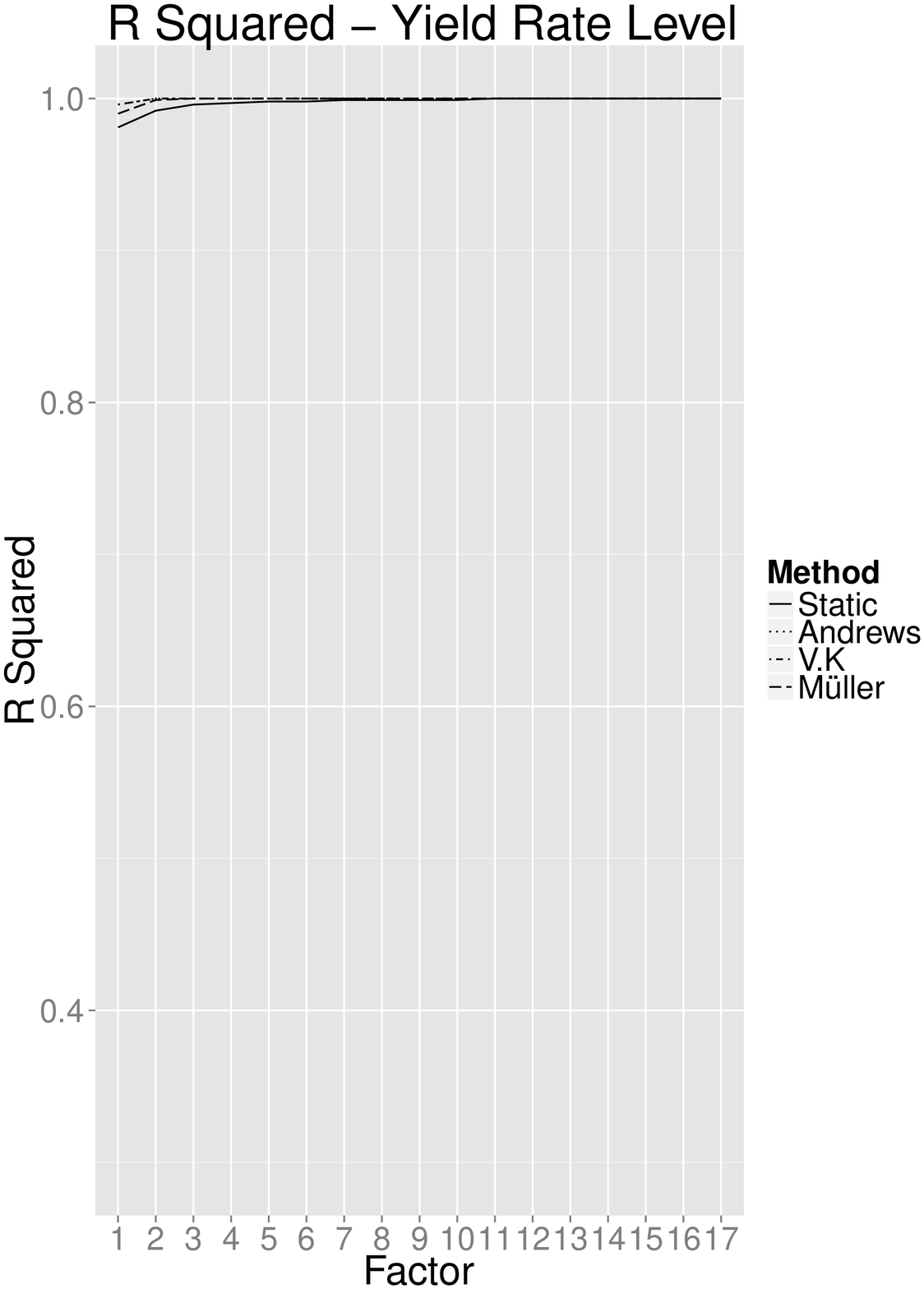}\includegraphics[width=6cm,height=5cm]{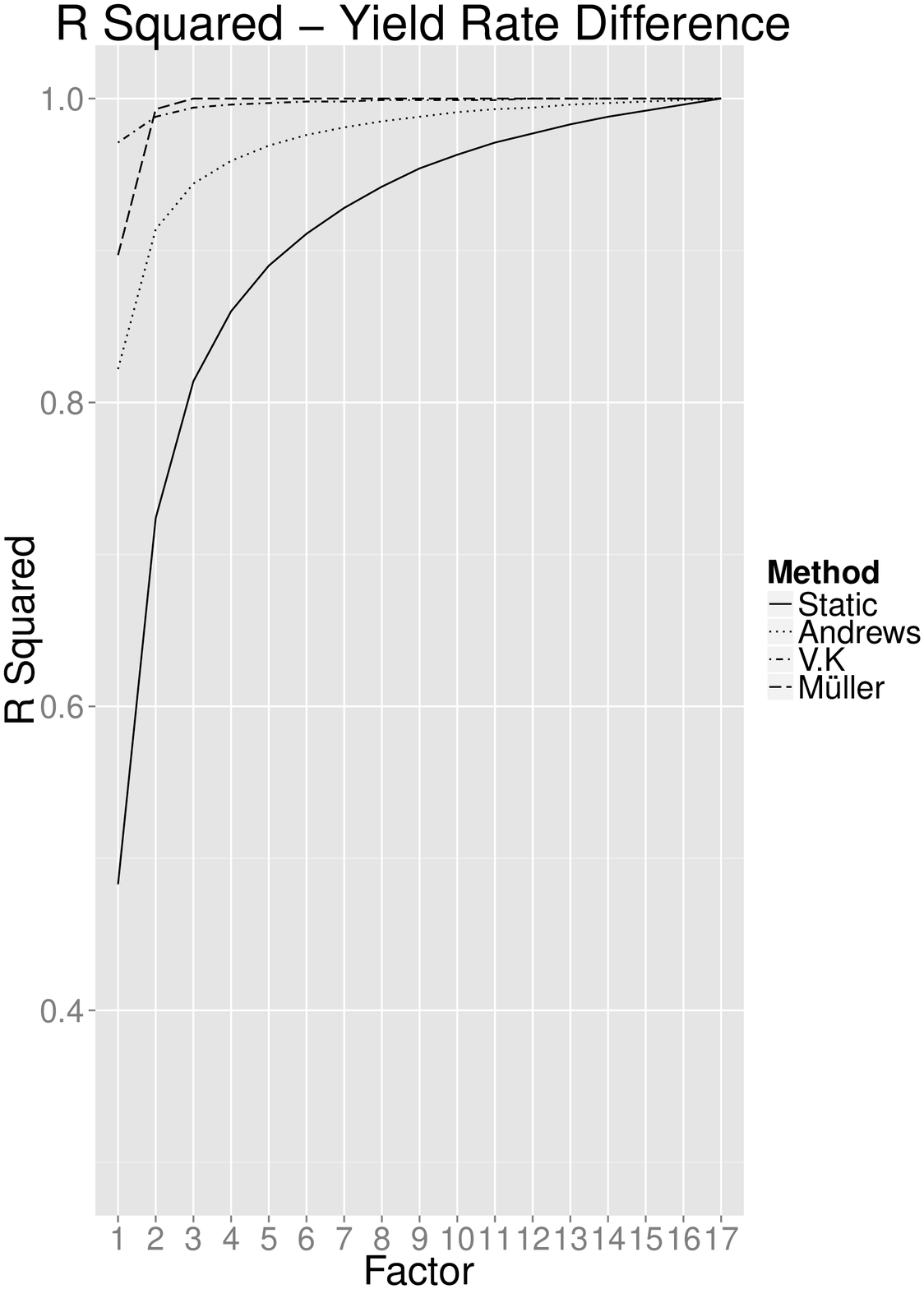}
\par\end{centering}

\caption{Cumulative $R^{2}$ obtained from the PCA decomposition for the level and first
difference of the forward and yield term structures. Second experiment with  observational errors generated by  market microstructure effects, Cox-Ingersoll-Ross process. Mean values from 1,000 Monte Carlo simulations.
\label{fig:Cox-Ingersoll-Ross} }
\end{figure}

We are going to study the impact of the MME as follows. In the sequel, the forward rate process $r$ in~(\ref{f1}) and the yield process $y$ in~(\ref{f2}) are given, respectively, by the Gaussian HJM and CIR models as specified above. Figure \ref{fig:Gaussian-HJM} reports the PCA methodology applied to

\begin{equation}\label{f1}
\left\{
\begin{array}{rl}
X_t(x) = r_t(x)+\varepsilon_t(x); & \hbox{where} \ \varepsilon~\text{is a Gaussian zero mean IID process with variance}~.0035 \\
Z_t(x) = y_t(x) + \eta_t(x) ;& \hbox{where} \ y~\text{is computed via formula}~(\ref{yld})~\text{and}~\eta_t(x) = \frac{1}{x}\int_0^x\varepsilon_t(r)dr.
\end{array}
\right.
\end{equation}
Figure \ref{fig:Cox-Ingersoll-Ross} reports the PCA methodology applied to

\begin{equation}\label{f2}
\left\{
\begin{array}{rl}
Z_t(x) = y_t(x)+\eta_t(x); & \hbox{where} \ \eta~\text{is a Gaussian zero mean IID process with variance}~.0035 \\
X_t(x) = r_t(x) + \varepsilon_t(x) ;& \hbox{where} \ \varepsilon_t(x) = \eta_t(x) + x\frac{\partial \eta_t}{\partial x}(x),~r_t(x) = y_t(x)+x\frac{\partial y_t}{\partial x}(x).
\end{array}
\right.
\end{equation}

We remarkably notice (see Figures \ref{fig:Gaussian-HJM} and \ref{fig:Cox-Ingersoll-Ross}) the same structure reported by the empirical analysis given in \citet{liu} when using the classical static covariance estimator $\hat{V}_{s}$ in the PCA methodology. We observe a substantial difference between the number of factors capable to explain the covariance structure of the yield and forward rate curves, specially when applied to the first-difference of the curves.

In Figures~\ref{fig:Gaussian-HJM} and \ref{fig:Cox-Ingersoll-Ross}, the estimates based on the standard sample covariance matrix $\hat{V}_s$ are highly biased for the forward rate curves. This can also be observed in the first-difference of the yield curves. We remarkable notice that both $\hat{V}_{lr}^{VK}$ and $\hat{V}_{lr}^{UA(p)}$ estimate the correct number of factors, especially when computed on the first-difference of the forward rate curves in which the contamination is more problematic. The $\hat{V}_{lr}^{A}$ estimator still indicate an excessive number of factors for the forward rate curves.

Figure~\ref{fig:Cox-Ingersoll-Ross} reports one typical situation of MME founded in practice. One should notice that the presence of additive IID Gaussian measurement errors in the observed yield curve does not affect the PCA estimation. Neither the usual sample static covariance matrix $\hat{V}_s$ nor $\hat{V}^{A}_{lr}, \hat{V}^{VK}_{lr}$ and $\hat{V}^{UA(p)}_{lr}$ are affected by the presence of this type of error in the yield curve. In contrast, the estimates based on $\hat{V}_s$ for the forward rate and first-difference of the yield curves are clearly biased. The situation is even worst when dealing with forward rate first-differences.

\begin{figure}
\begin{centering}
\includegraphics[width=6cm,height=5cm]{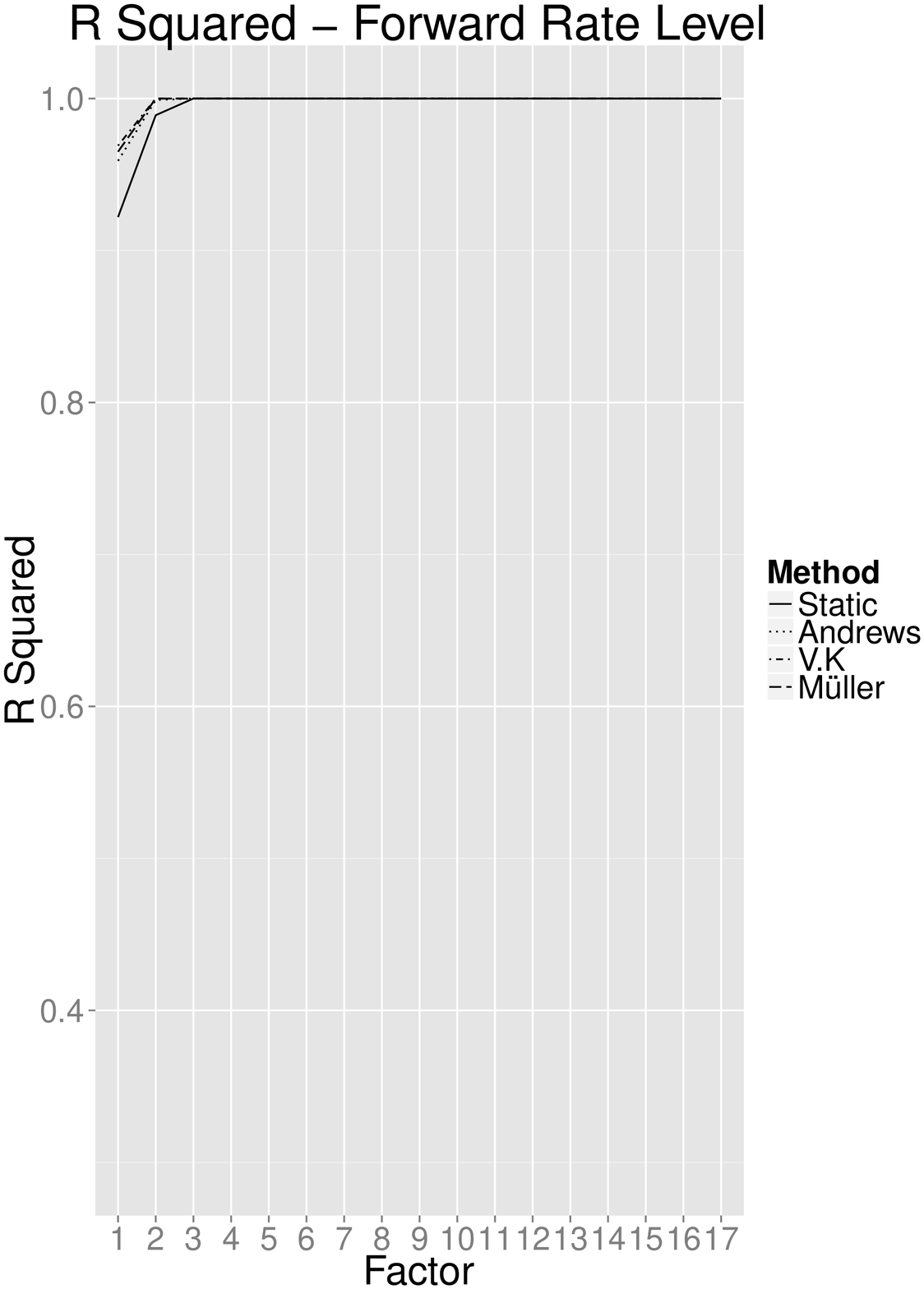}\includegraphics[width=6cm,height=5cm]{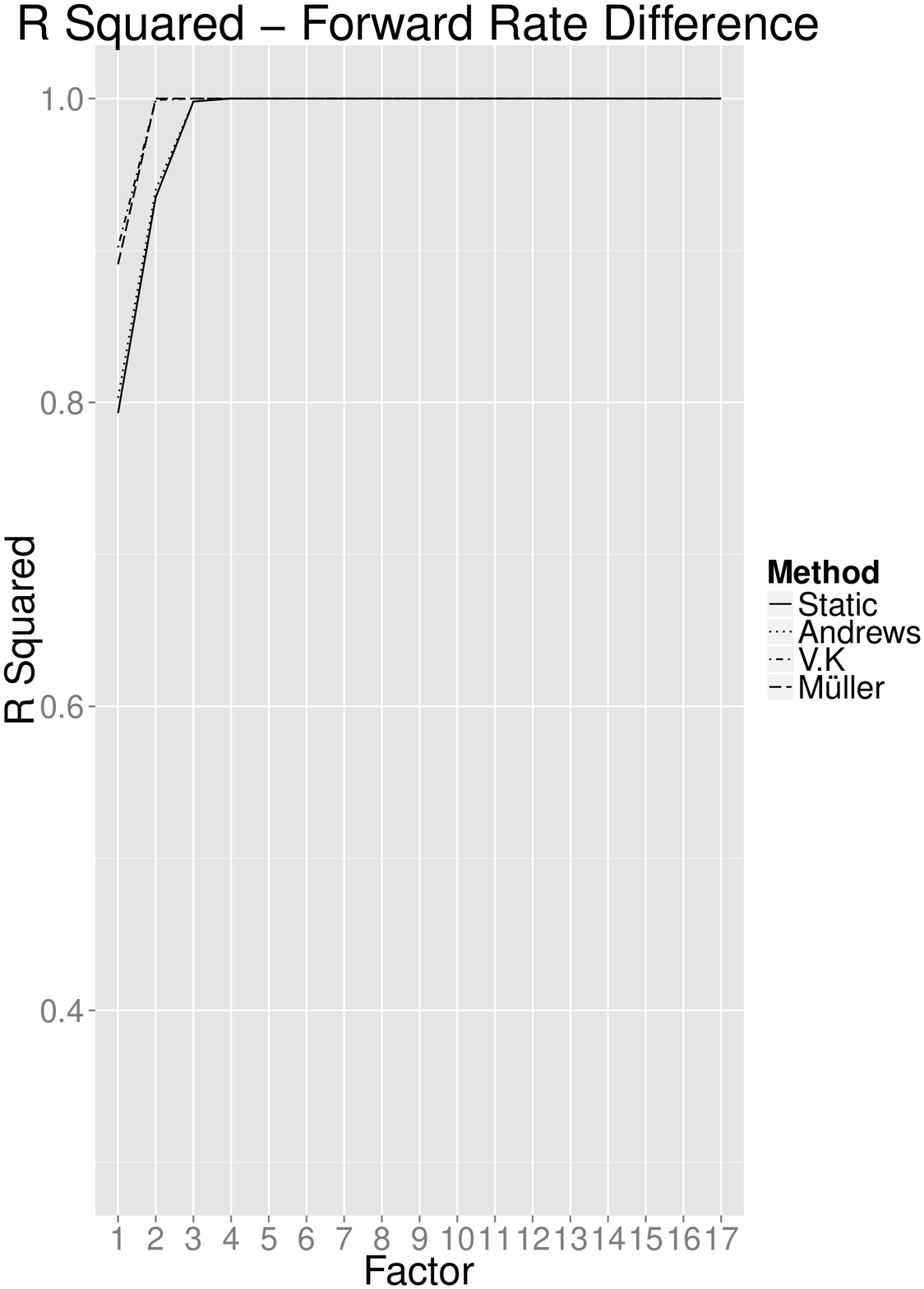}
\par\end{centering}

\begin{centering}
\includegraphics[width=6cm,height=5cm]{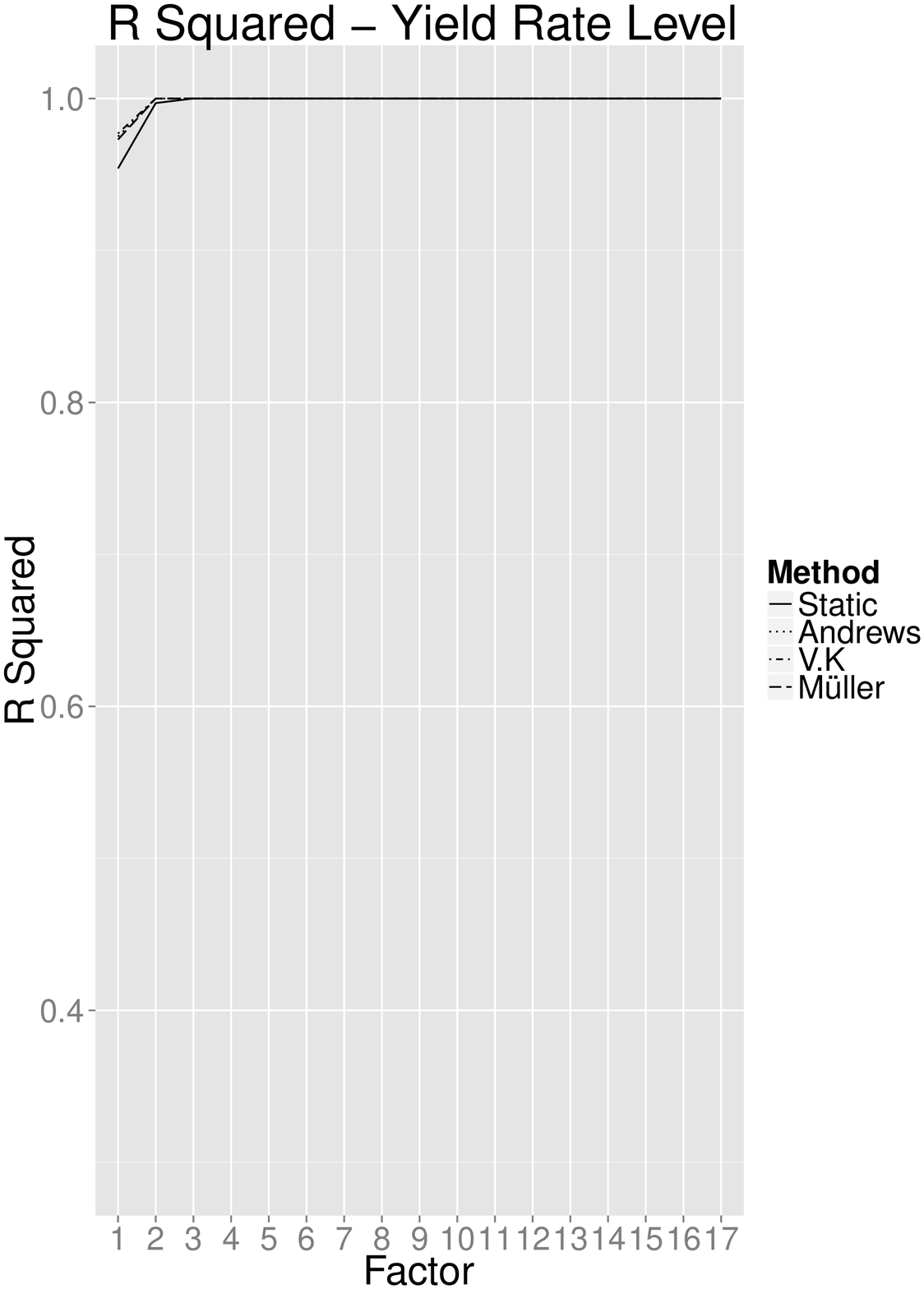}\includegraphics[width=6cm,height=5cm]{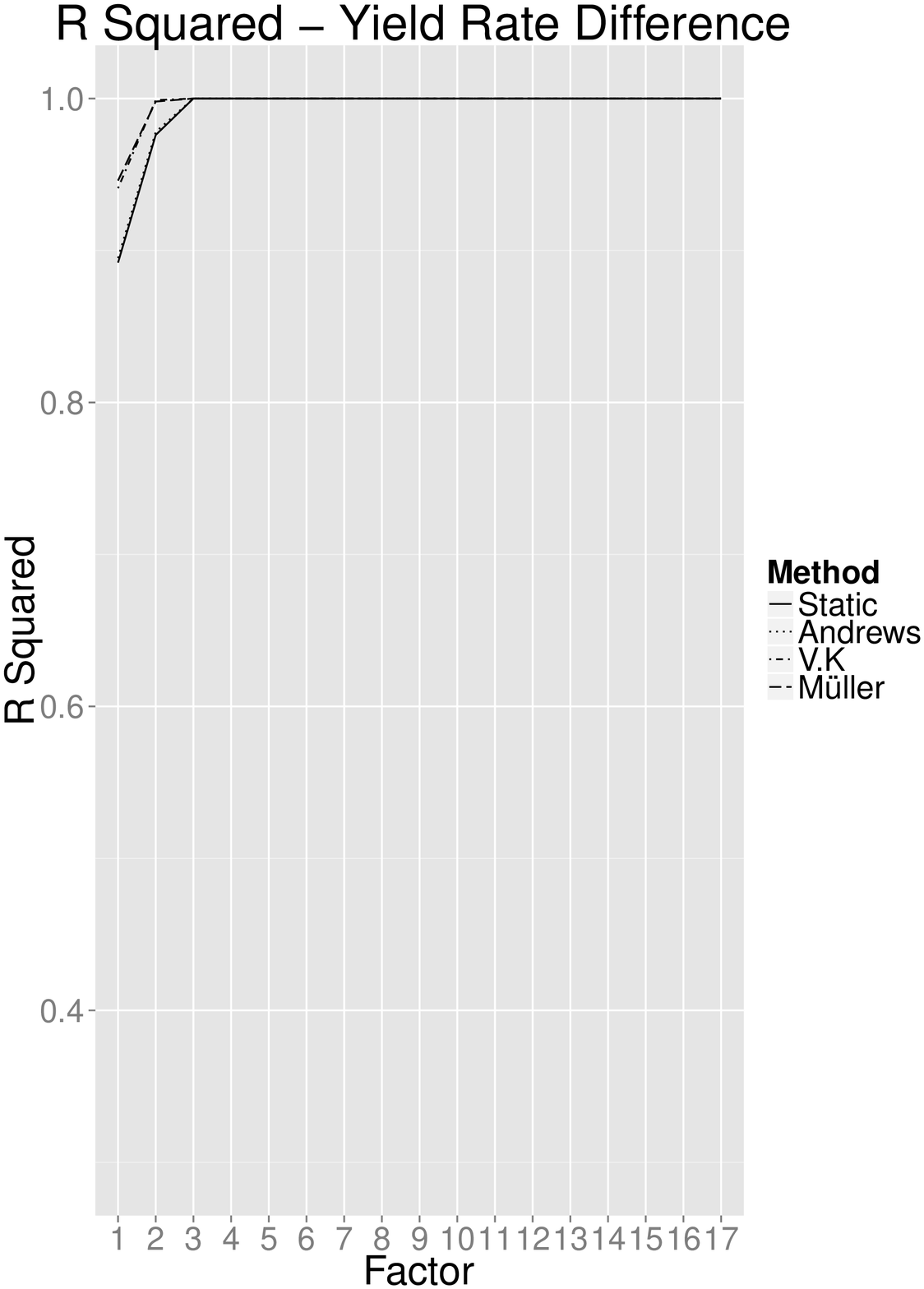}
\par\end{centering}

\begin{centering}
\caption{Cumulative $R^{2}$ obtained from the PCA decomposition for the level and first
difference of the forward and yield term structures. First experiment without observational errors, Gaussian HJM process\label{fig:zeroGaussian-HJM}. Mean values from
1,000 Monte Carlo simulations.}

\par\end{centering}

\end{figure}

\begin{figure}
\begin{centering}
\includegraphics[width=6cm,height=5cm]{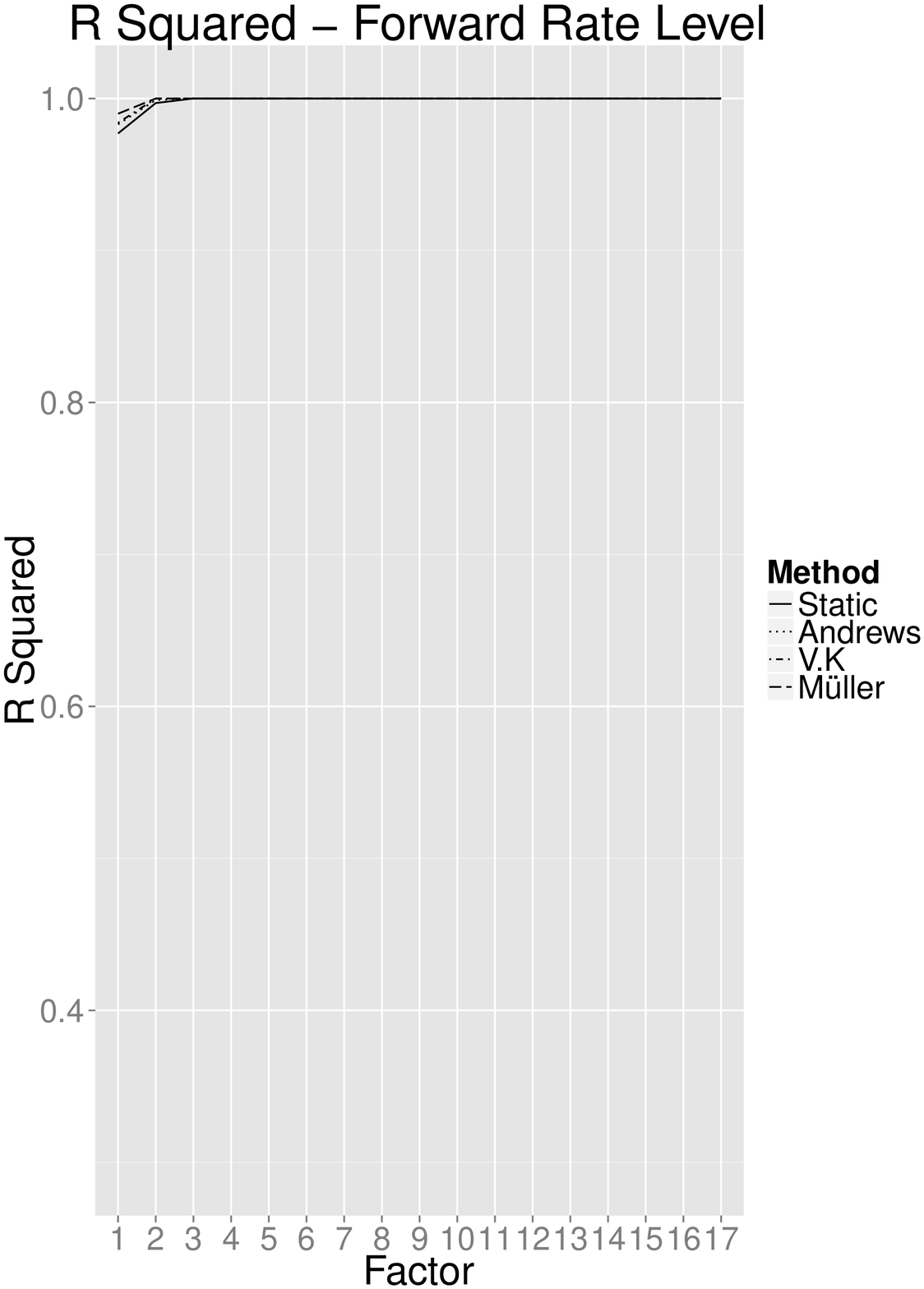}\includegraphics[width=6cm,height=5cm]{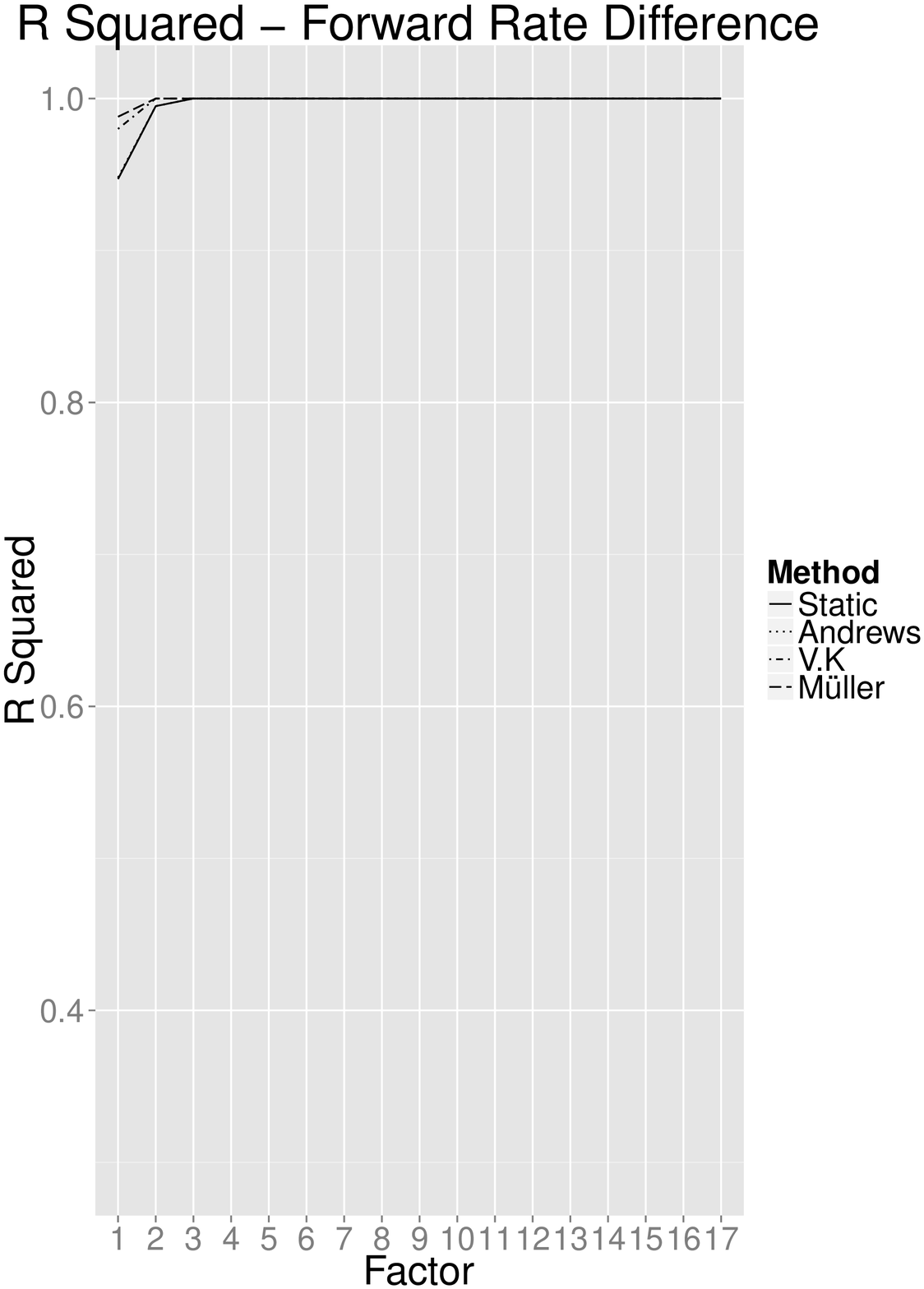}
\par\end{centering}

\begin{centering}
\includegraphics[width=6cm,height=5cm]{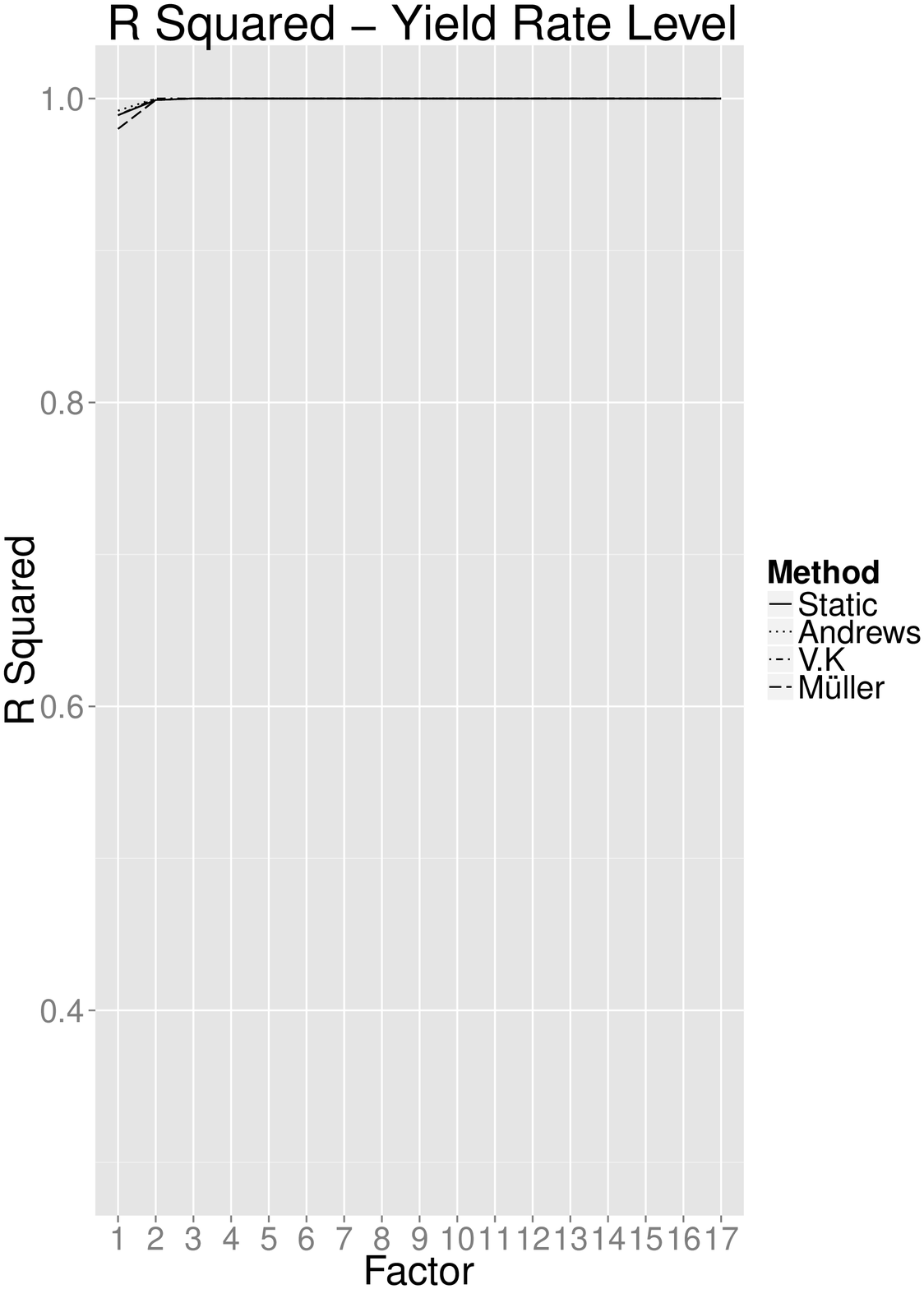}\includegraphics[width=6cm,height=5cm]{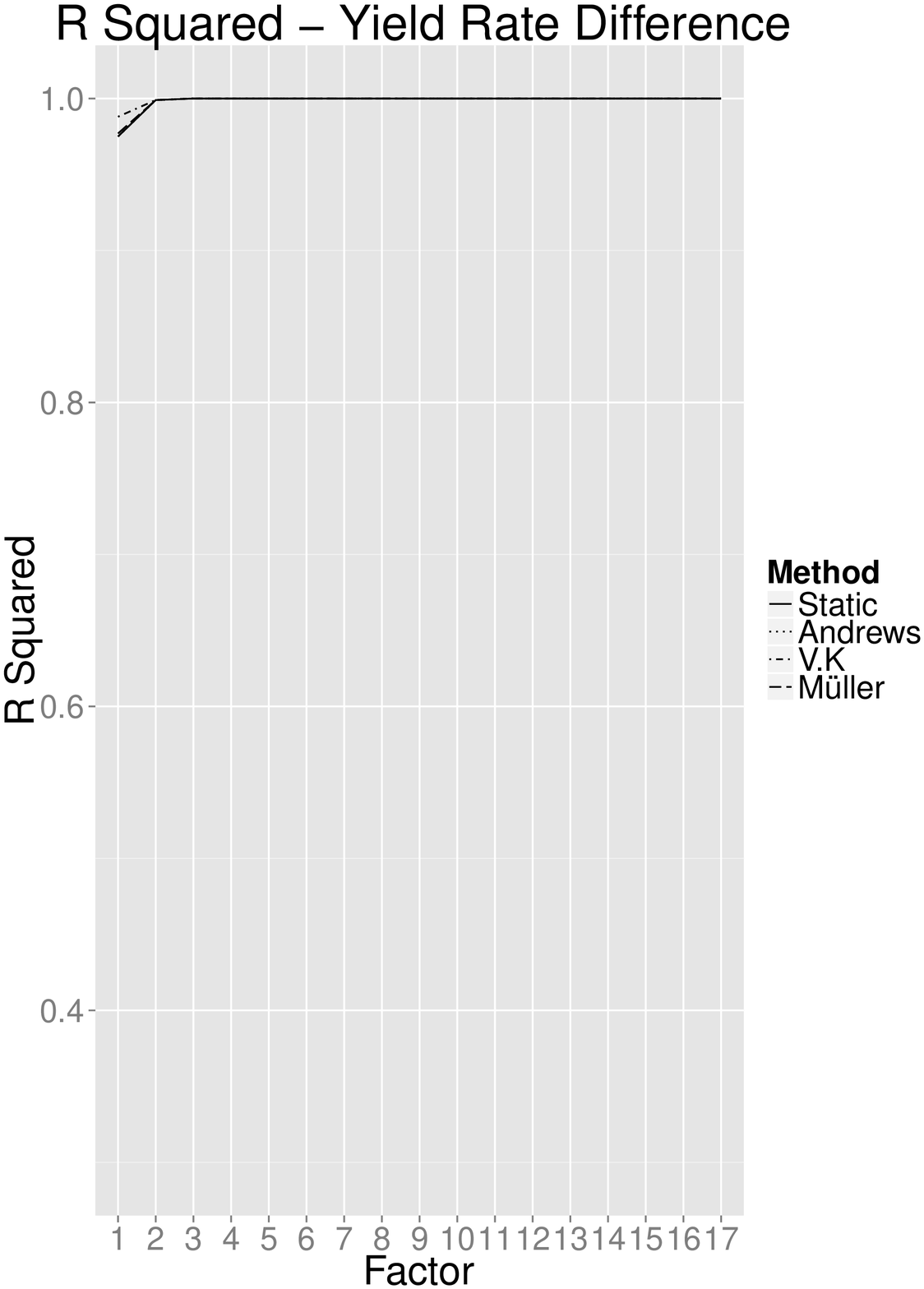}
\par\end{centering}

\caption{Cumulative $R^{2}$ obtained from the PCA decomposition for the level and first
difference of the forward and yield term structures . Second experiment without observational errors, Cox-Ingersoll-Ross process. Mean values from 1,000 Monte Carlo simulations.
\label{fig:zeroCox-Ingersoll-Ross} }
\end{figure}

The explanation for the results reported in Figures \ref{fig:Gaussian-HJM} and \ref{fig:Cox-Ingersoll-Ross} lies in the following argument. We readily see the existence of MME in bond markets impacts differently forward rate and yield curves. In~(\ref{f2}), the MME introduces a moving average structure with negative persistence into the forward rate curve due to the components $\frac{\partial y_t}{\partial x}$ and $\frac{\partial \eta_t}{\partial x}(x)$. One should notice that the existence of a moving average error structure affects severally the usual estimators of the covariance matrix. See \citet{key-4} and \cite{ECTover} for a discussion about this issue. In contrast, the MME in~(\ref{f1}) only introduces an IID component into the observed yield curve process. This IID component does not affect the temporal dependence of the yield curve. The results reported in Figures \ref{fig:zeroGaussian-HJM}-\ref{fig:Cox-Ingersoll-Ross} show that the PCA methodology applied to forward rate curves is a very delicate issue because it might be subject to non-negligible observational errors due to MME. The presence of MME explains the large difference between the number of components indicated by the PCA method for yield and forward curves. This remark is particularly more evident when double differentiation of the underlying moving average process appears due to the first-difference of the observed forward rate curves in~(\ref{f2}). The main reason to use LRCM estimators instead of the usual $\hat{V}_s$ is to handle these types of observational errors. In particular, our results strongly indicate the estimators $\hat{V}^K_{lr}$ and $\hat{V}^{UA(p)}_{lr}$ have successfully filtered the MME in the PCA methodology.

\begin{figure}
\begin{centering}
\includegraphics[width=6cm,height=5cm]{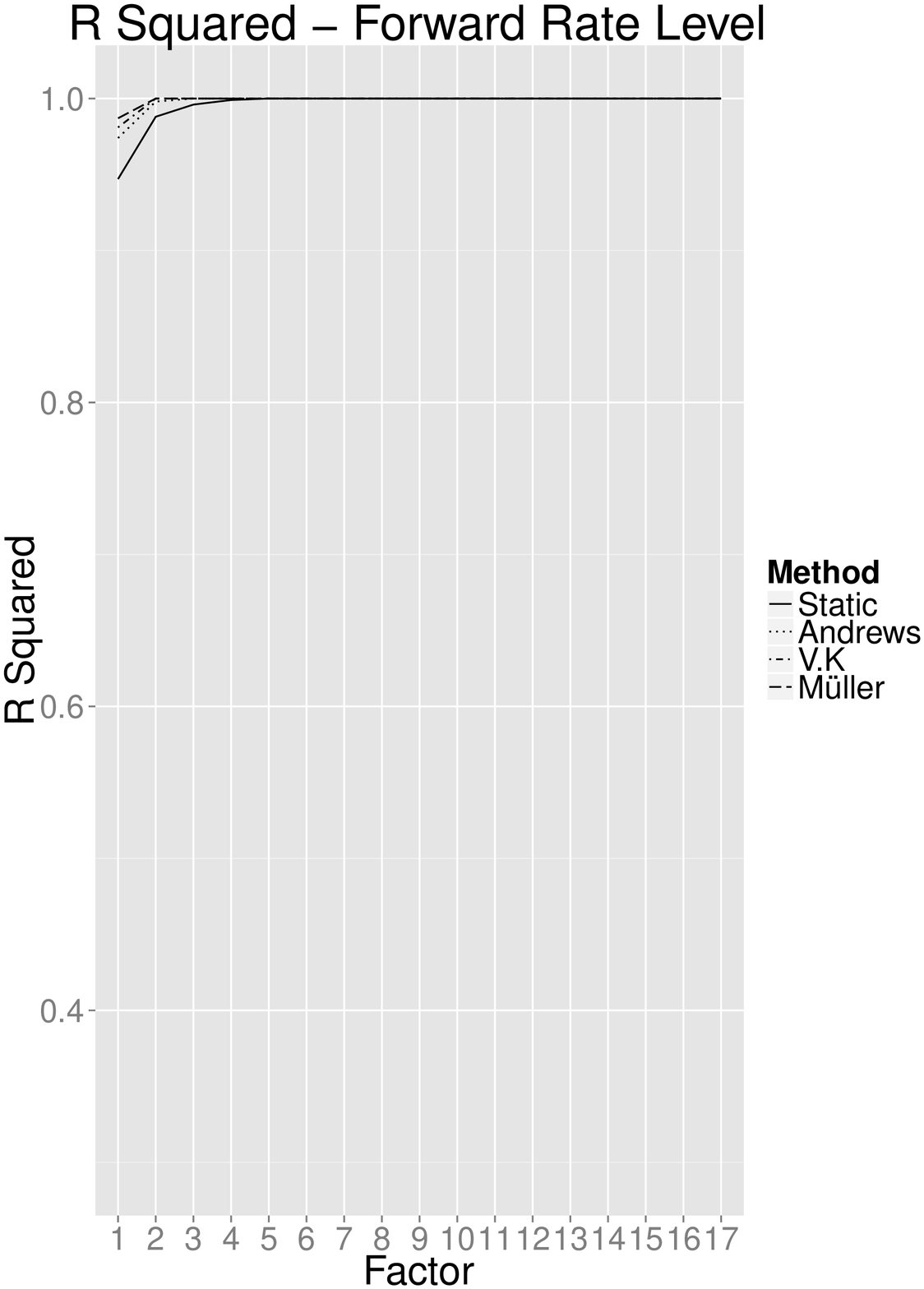}\includegraphics[width=6cm,height=5cm]{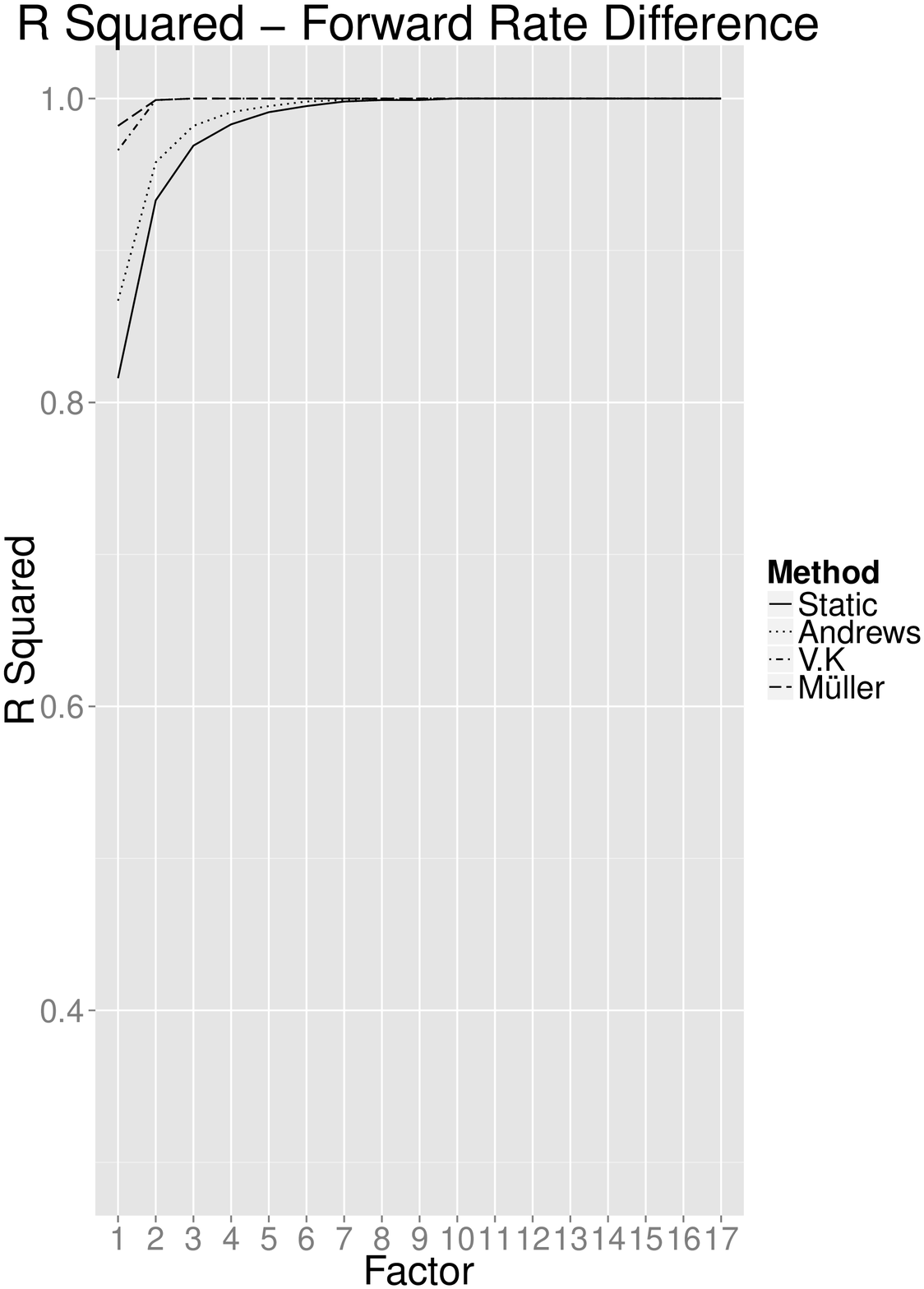}
\par\end{centering}

\begin{centering}
\includegraphics[width=6cm,height=5cm]{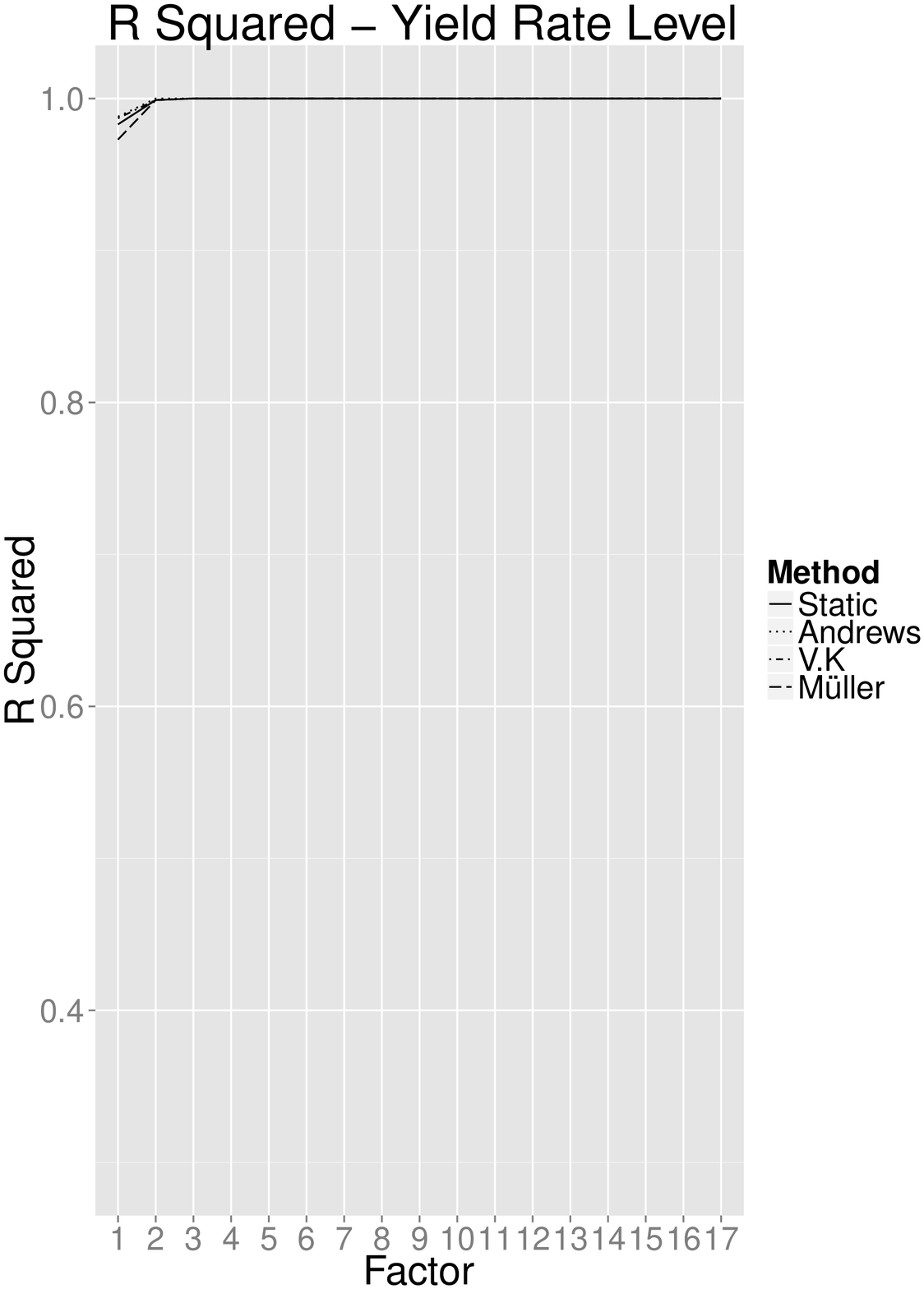}\includegraphics[width=6cm,height=5cm]{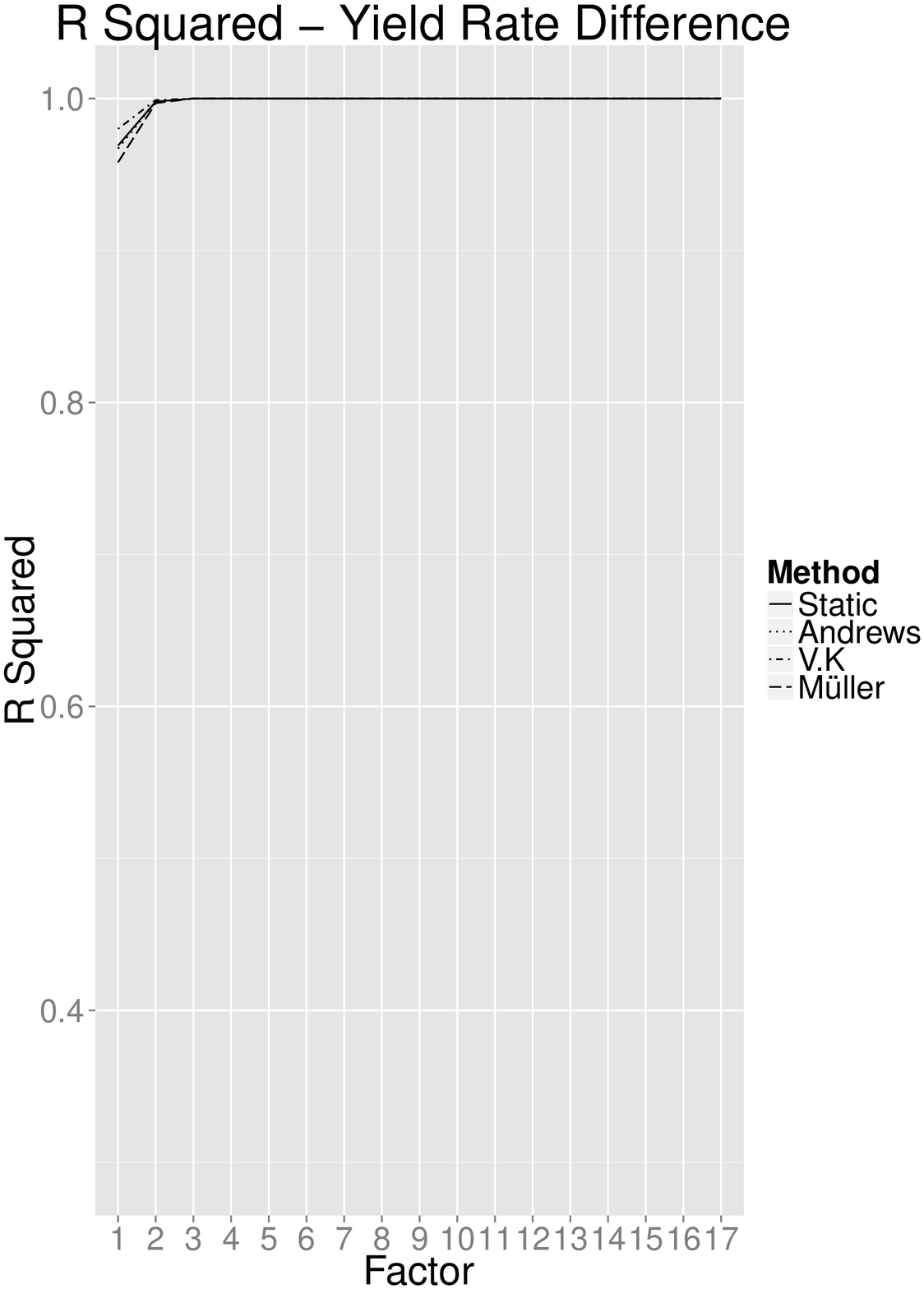}
\par\end{centering}

\caption{Cumulative $R^{2}$ obtained from the PCA decomposition for the level and first
difference of the forward and yield term structures. Experiment with observational errors generated by  cubic spline interpolation, Cox-Ingersoll-Ross process. Mean values from 1,000 Monte Carlo simulations.
\label{fig:splineCox-Ingersoll-Ross} }
\end{figure}

Let us now treat the case of a non-additive error structure $G(y_t,\eta_t)$ due to interpolation of the yield curve $y_t$ as described in \textbf{(A-B)}. Figure \ref{fig:splineCox-Ingersoll-Ross} reports the PCA methodology applied to

\begin{equation}\label{f3}
\left\{
\begin{array}{rl}
Z_t(x) = G(y_t(x),\eta_t(x)); & \hbox{where} \ G~\text{and}~\eta~\text{are generated by an IES induced by a cubic splines} \\
X_t(x) = Z_t(x) + x\frac{\partial Z_t}{\partial x}(x), &
\end{array}
\right.
\end{equation}
where $y$ is given by the CIR model as specified above. In this analysis, the error is generated by the classical spline method of \cite{mcculloch}. We simulate bond prices generated by the CIR model and we omit prices for 4 maturities randomly chosen, generating again 1,000 replications for this experiment. The omitted prices are then interpolated by cubic spline using the price curve, similar to the method of \cite{mcculloch}. Hence, observational errors for bond prices are then generated by means of interpolation. From these prices, we generate yield and forward rate interpolated curves. The PCA method based on the estimators of  Section~\ref{methods} is then applied to $(Z,X,\tilde{Z},\tilde{X})$ as specified in~(\ref{f3}).

In Figure \ref{fig:splineCox-Ingersoll-Ross}, the observational errors generated by the cubic splines do not significantly affect the number of factors in yield curves neither in level nor first-difference. However, it has again a nontrivial impact on the forward rate first-difference $\tilde{X}$. The number of factors of $\tilde{X}$ computed by the PCA based on $\hat{V}_{s}$ and $\hat{V}_{lr}^{A}$ is not correctly estimated. The standard PCA method based on $\hat{V}_s$ indicates the incorrect number of six factors which explain 99 \% of total variation. Similar to the analysis of MME given in Figures \ref{fig:Gaussian-HJM} and \ref{fig:Cox-Ingersoll-Ross}, the PCA based on $\hat{V}_{lr}^{VK}$ and $\hat{V}_{lr}^{UA(p)}$ again correctly estimates the true number of three factors.

The results obtained from these experiments show that the presence of observational errors generated by MME or IES induce a significant bias in the classical PCA methodology applied to forward rate curves. More importantly, our analysis shows that the LRCM estimators ($\hat{V}_{lr}^{VK}, \hat{V}_{lr}^{UA(p)}$) indicate the correct number of factors in the Monte Carlo experiments. The results of this section are robust w.r.t to model specification and types of observational errors.

%As shown the result of the difference between the number of factors the forward curve and spot curve depends crucially on the contamination by measurement errors, in practice generated by the effects of market microstructure or observational errors introduced by the procedure of construction of the forward curve.

\section{Empirical analysis for the numbers of factors}\label{empan}
In this section, we compare the results described in Section~\ref{numericalres} with a PCA application to two distinct real data sets. The first one is given by Treasury bonds (zero coupon) with maturities 3, 6, 9, 12, 15, 18, 21, 24, 30, 36, 48, 60, 72, 90, 108 and 120 months (17 maturities), with monthly observations ranging from 1985-Jan to 2000-Dez. This data set is constructed based on the Fama-Bliss methodology (unsmoothed
Fama-Bliss) and it was already used by \citet{key-12}. The second data set is the UK term-structure obtained from the Bank of England with maturities .5 to 25 years (50 maturities) and daily data ranging from Jan/4/2005 to Feb/29/2012 summing 1872 observations.

\begin{figure}
\begin{centering}
\includegraphics[width=6cm,height=5cm]{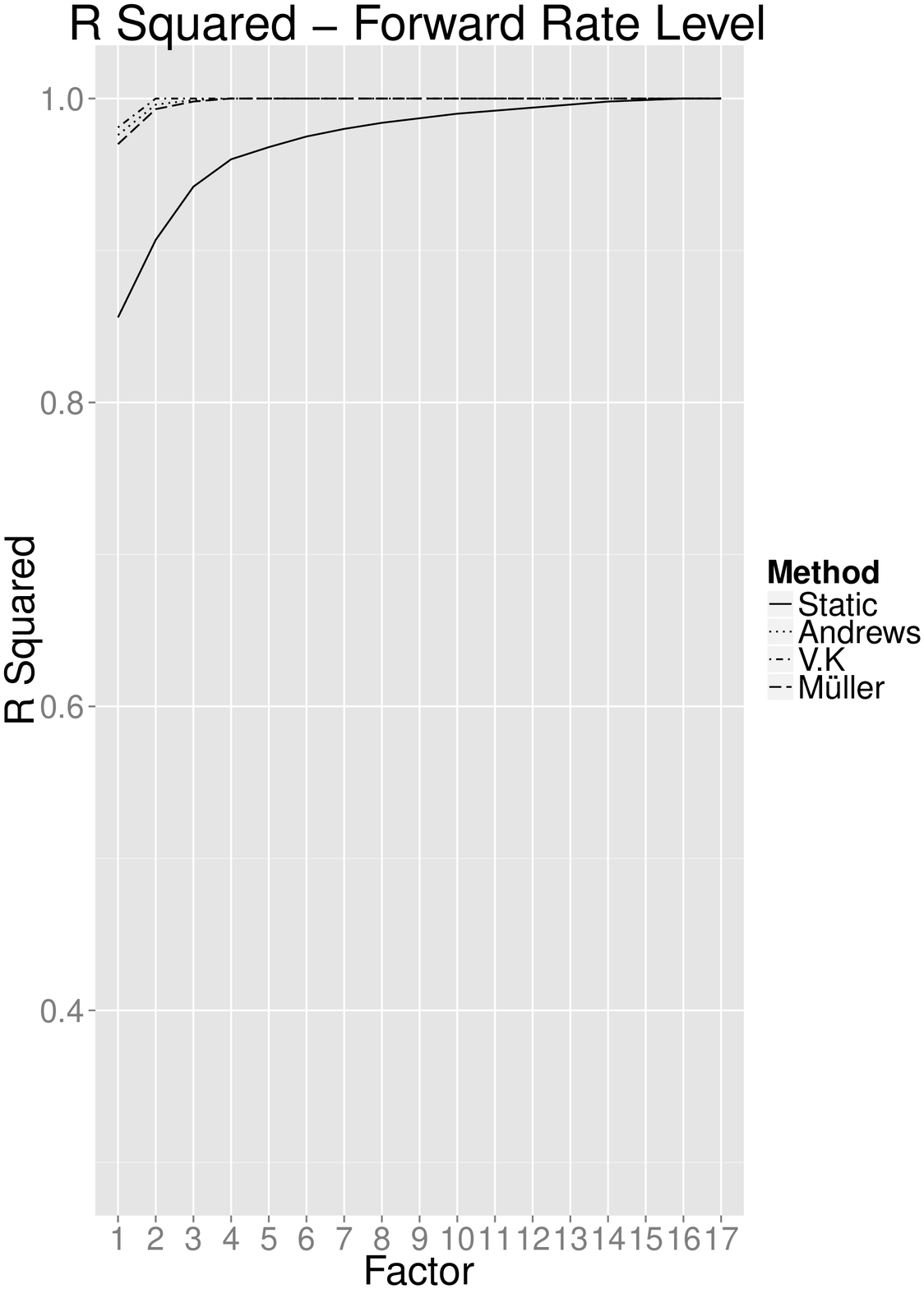}\includegraphics[width=6cm,height=5cm]{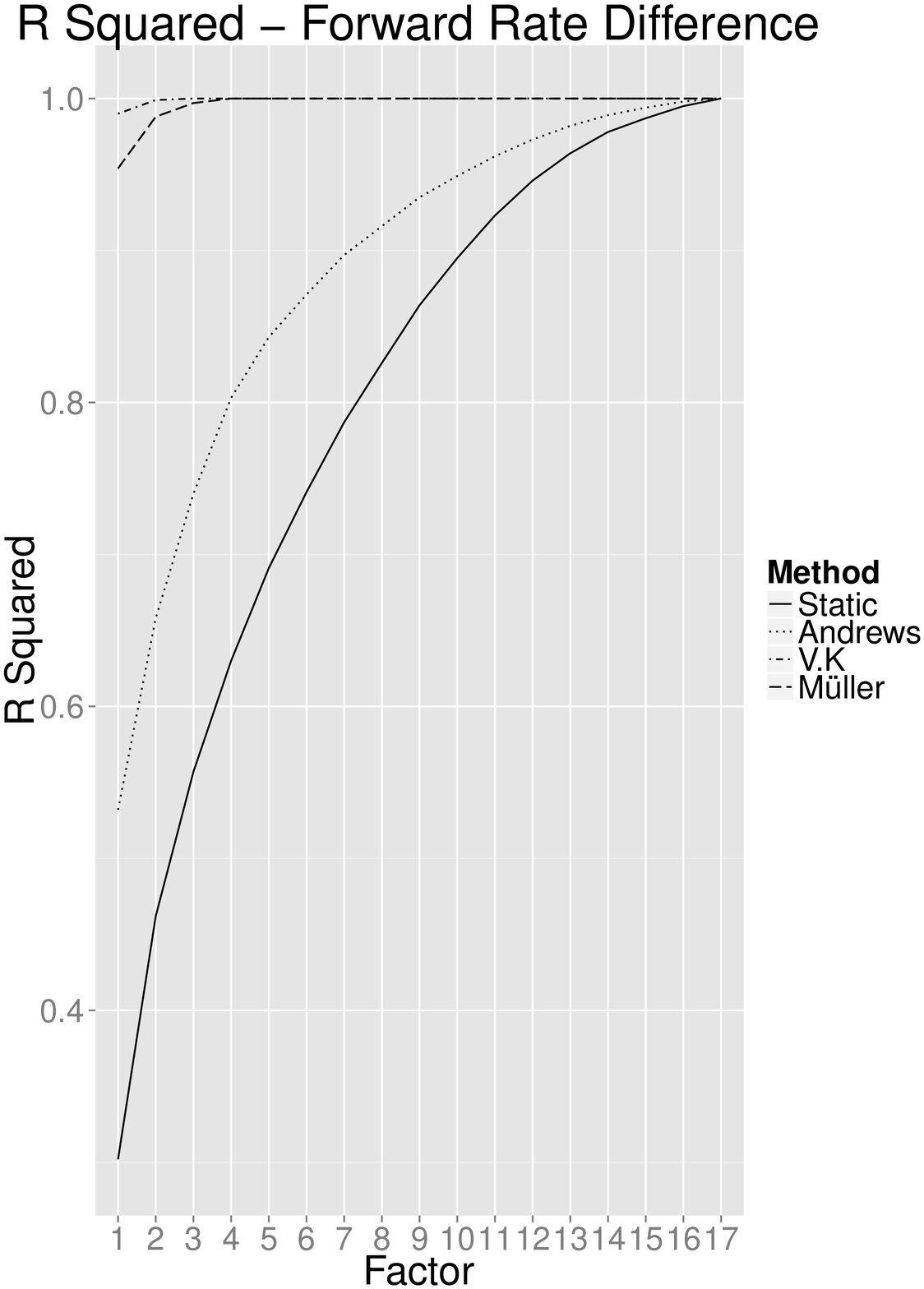}
\par\end{centering}

\begin{centering}
\includegraphics[width=6cm,height=5cm]{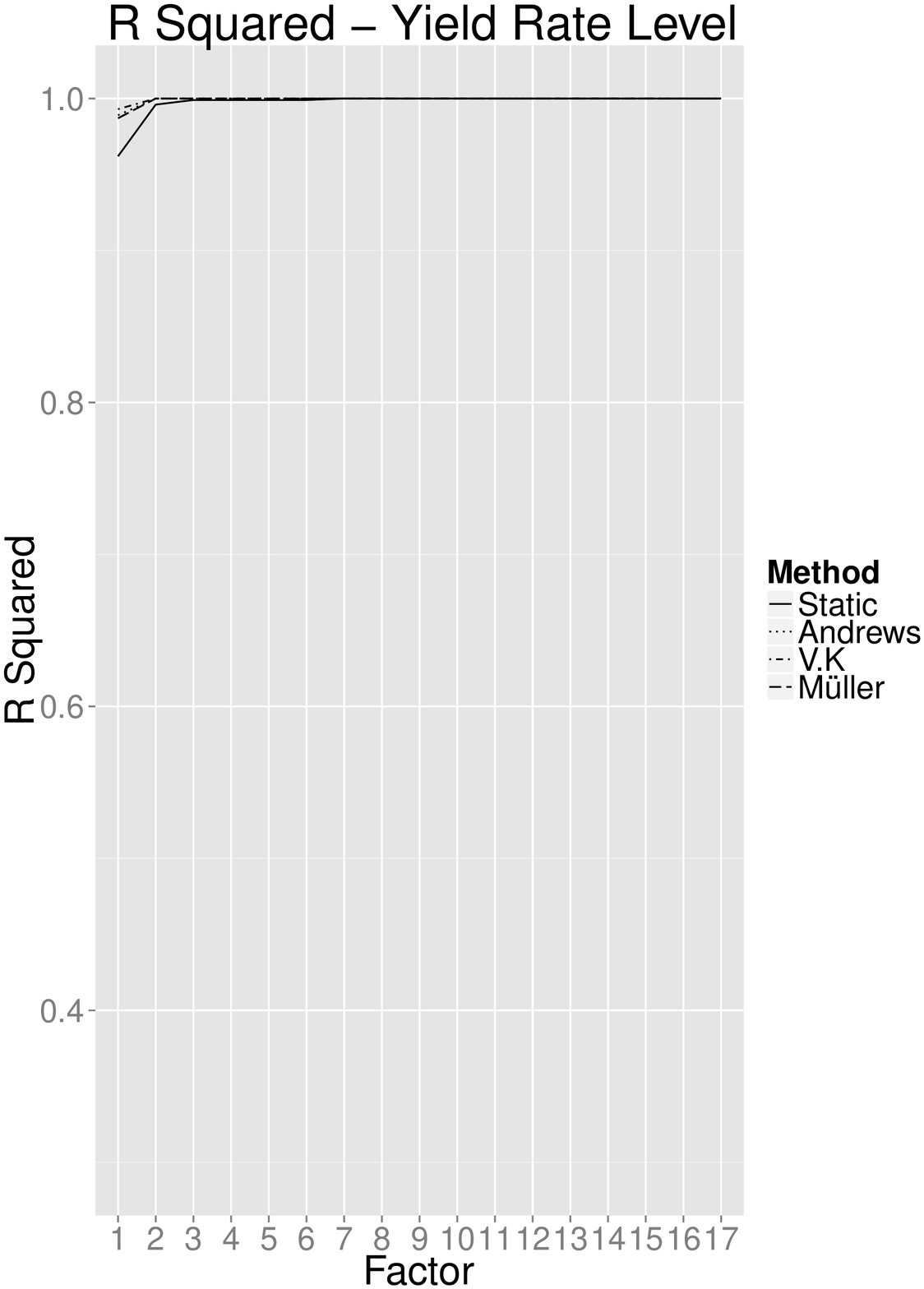}\includegraphics[width=6cm,height=5cm]{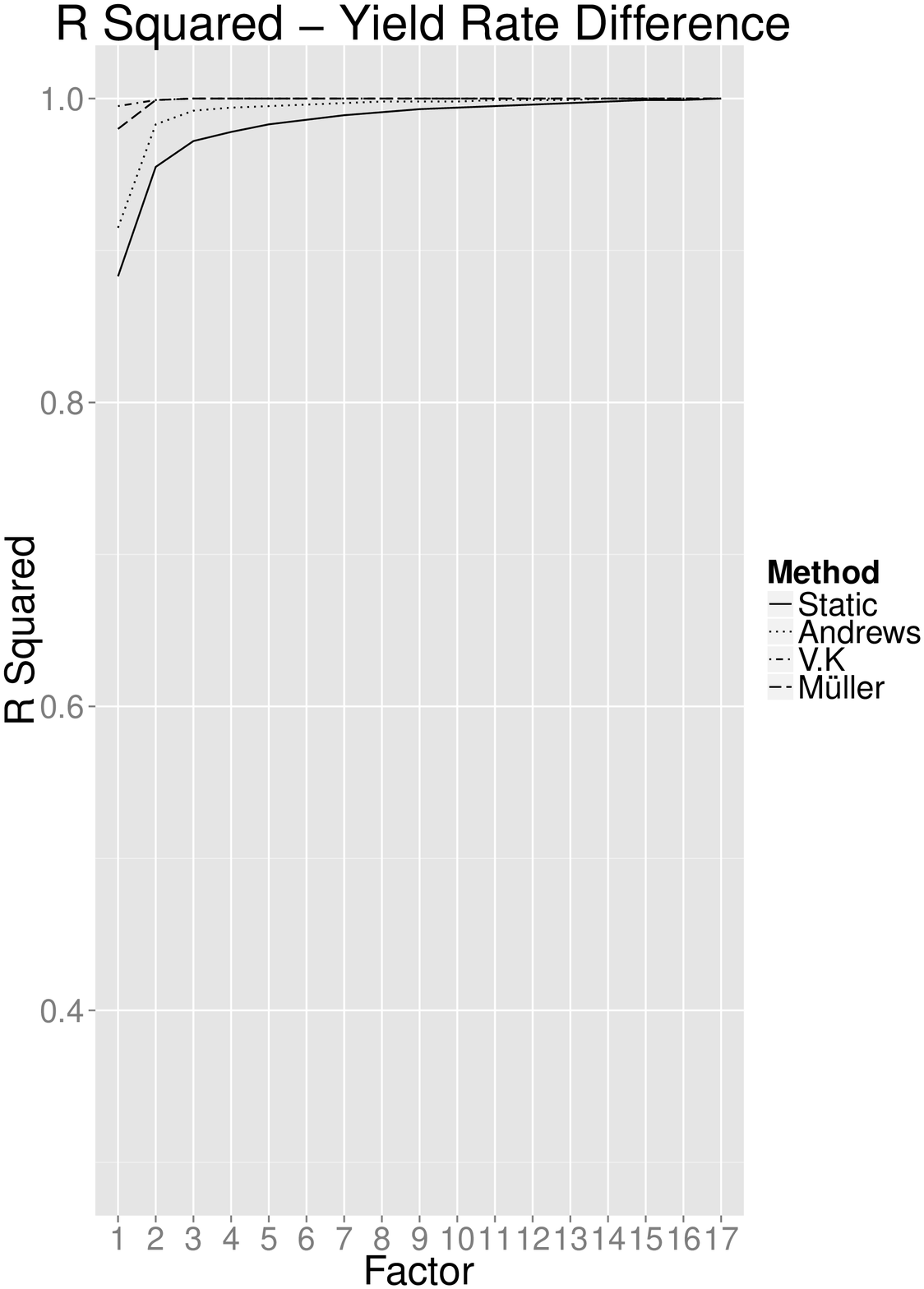}
\par\end{centering}

\centering{}\caption{Cumulative $R^{2}$ obtained from the PCA decomposition for the level and first
difference of the forward and yield term structures - U.S. Treasury
Curve - Fama-Bliss Database.\label{fig:U.S.-Treasury-Curve}}
\end{figure}

\begin{figure}
\begin{centering}
\includegraphics[width=6cm,height=5cm]{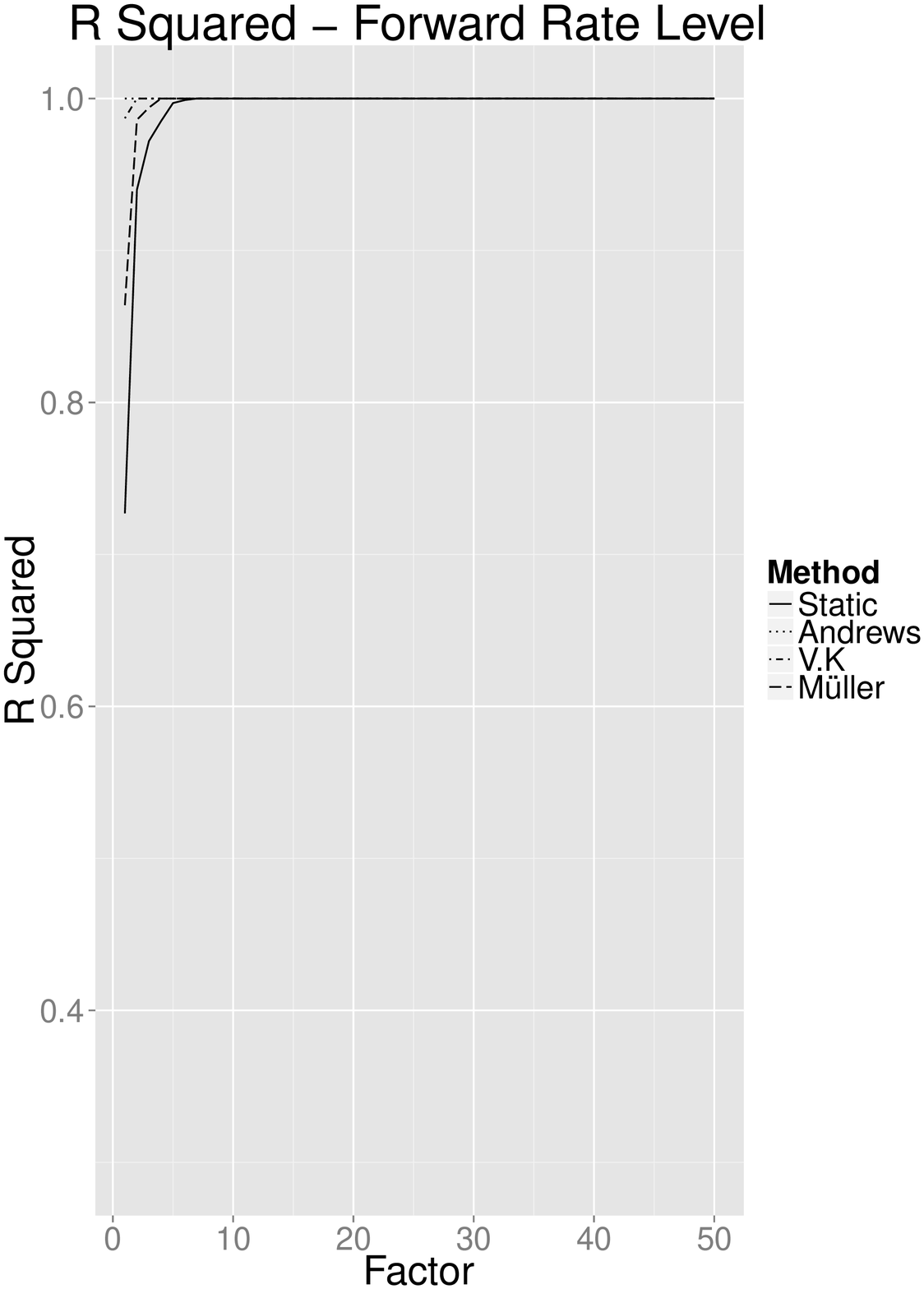}\includegraphics[width=6cm,height=5cm]{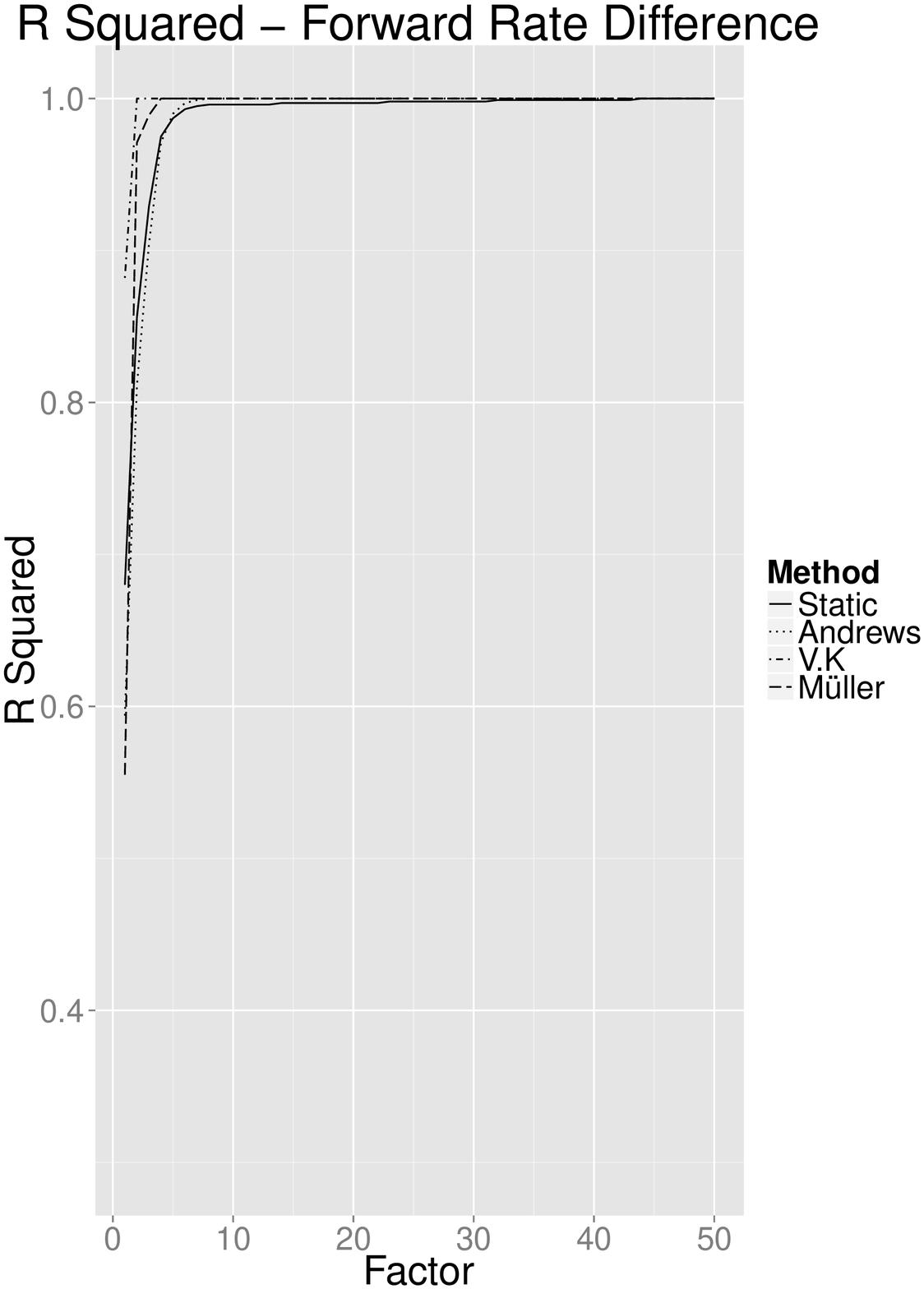}
\par\end{centering}

\begin{centering}
\includegraphics[width=6cm,height=5cm]{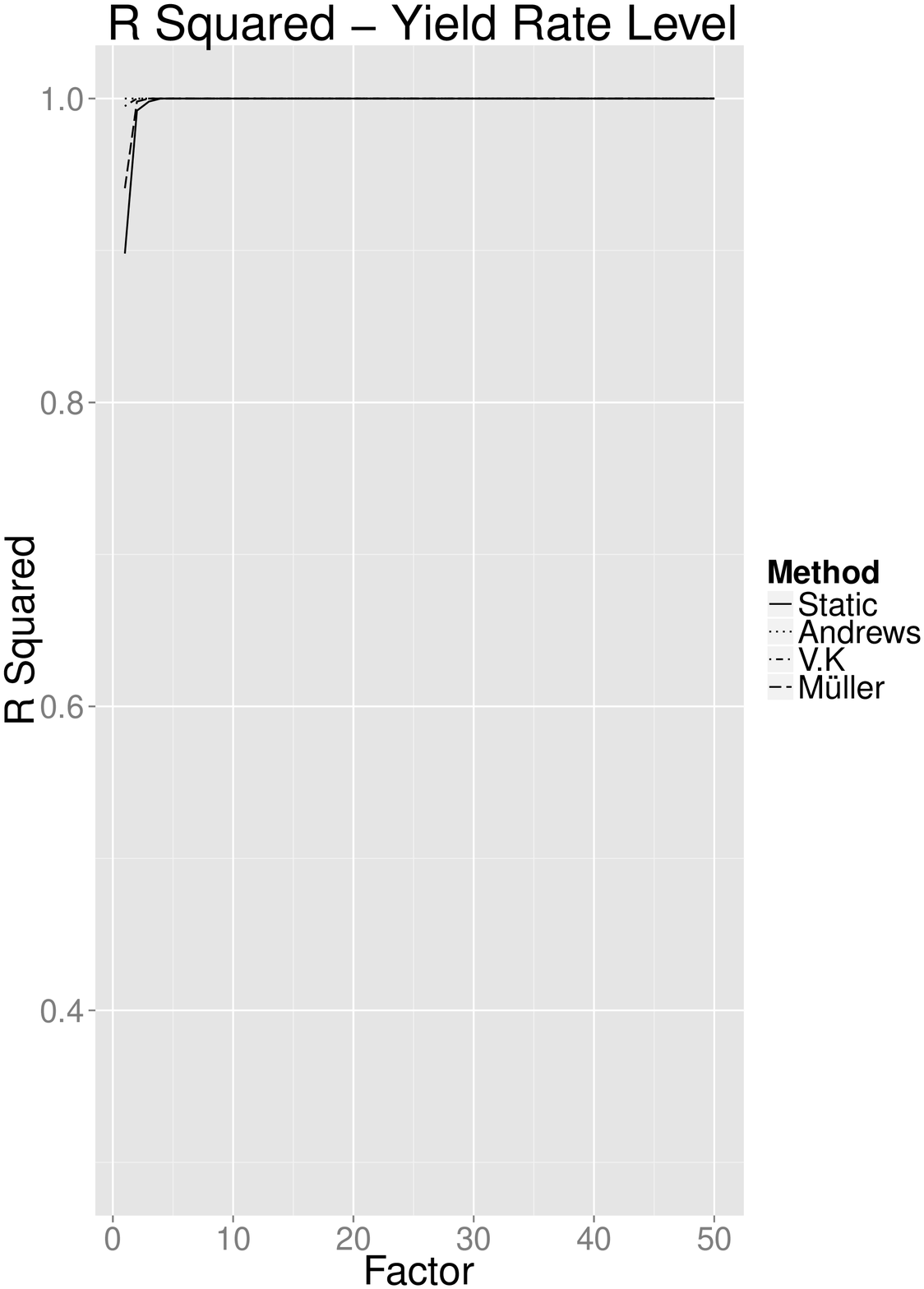}\includegraphics[width=6cm,height=5cm]{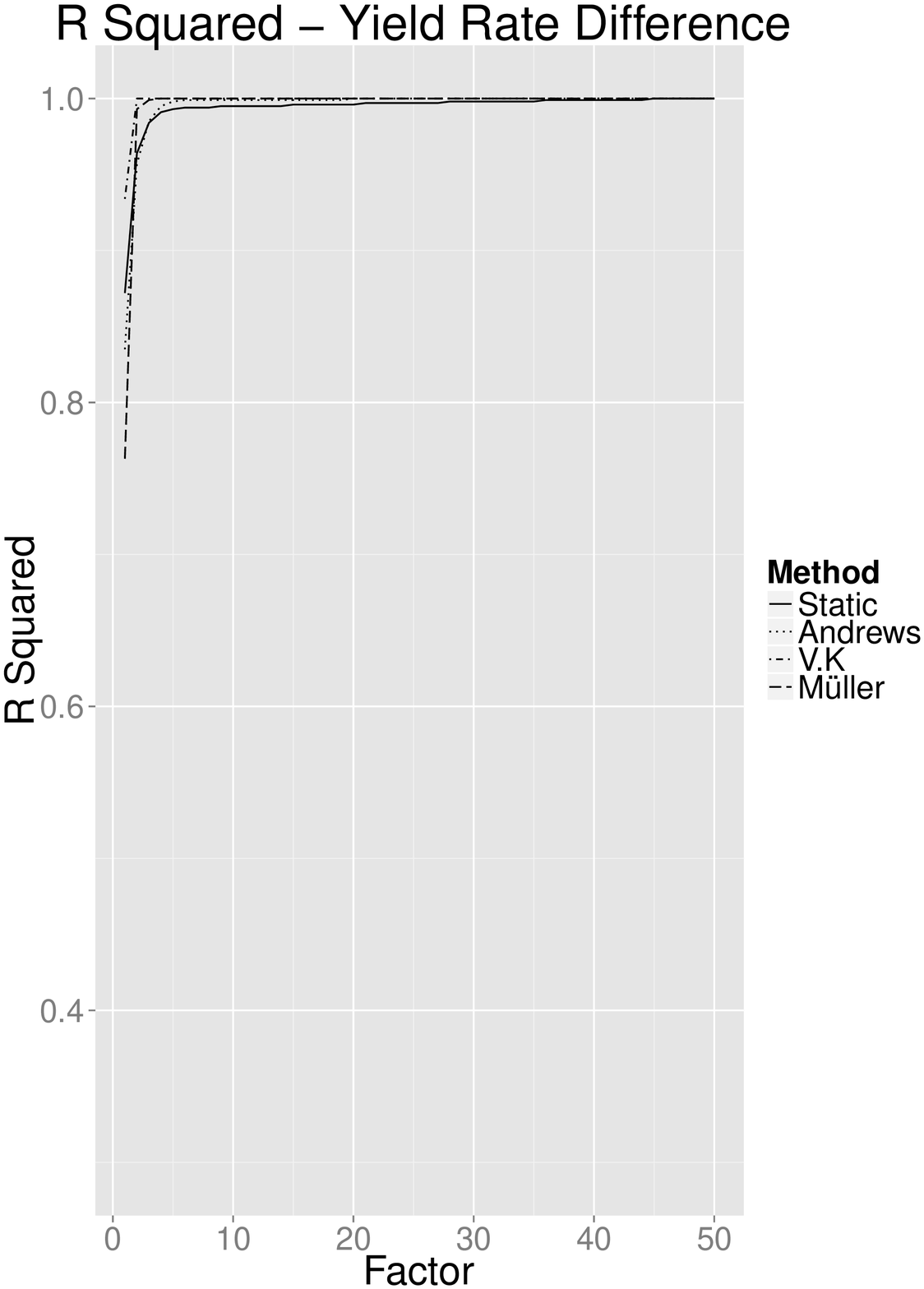}
\par\end{centering}

\centering{}\caption{Cumulative $R^{2}$ obtained from the PCA decomposition for the level and first
difference of the forward and yield term structures - U.K. Bank of England Term Structure Database.\label{fig:U.K.-Treasury-Curve}}
\end{figure}

In Figures \ref{fig:U.S.-Treasury-Curve} and \ref{fig:U.K.-Treasury-Curve}, we report the cumulative $R^{2}$ obtained from the PCA decomposition for the forward rate and yield curves of the U. S. and U. K. markets, respectively. The usual static covariance matrix estimator $\hat{V}_{s}$ applied to these two data sets shows the same behavior as described by \citet{liu}. We observe a large difference for the number of principal components between yield and forward rate curves, both in the first-difference and level. In particular, we observe the same type of behavior as described in Section~\ref{numericalres}. More importantly, we notice that both estimators $\hat{V}_{lr}^{VK}$ and $\hat{V}_{lr}^{UA(p)}$ indicate the same number of factors between yield and forward rate curves. This result holds for both the first-difference and the level of the curves. More importantly, it corroborates with our findings in Figure \ref{fig:Cox-Ingersoll-Ross}.

%According to Proposition \ref{numberres}, these results strongly suggest the presence of contamination errors in the data. The empirical applications reproduce the same effects observed in Section~\ref{numericalres}. More precisely, we observe the presence of observational errors generating a large divergence in the number of factors by using the standard PCA method on static covariance matrix. See Figure

According to Proposition \ref{numberres}, these results strongly suggest the presence of contamination errors in the data. In the US data, the error contamination might be explained by the Fama-Bliss method of construction of the forward rate curve (\cite{fama}) which uses a piecewise constant function to approximate the discount factor. The Bank of England term structure is constructed by means of a smooth cubic spline interpolation method for the yield and forward rate curves (see e.g~\cite{anderson} for details). Therefore, it is subject to observational errors similar to the Monte Carlo experiment ~(\ref{f3}) in Figure \ref{fig:splineCox-Ingersoll-Ross}. The presence of MME in the US data might also be a source of observational errors as reported by~\cite{JFQ} and~\cite{Mizrach}. The reader is urged to compare the Monte Carlo experiments~(\ref{f2}) and (\ref{f3}) to Figures \ref{fig:U.S.-Treasury-Curve} and \ref{fig:U.K.-Treasury-Curve}, respectively. The empirical analysis of this section jointly with Section~\ref{numericalres} and Proposition~\ref{numberres} give strong support for the use of PCA based on the LRCM estimators described in Section \ref{methods}.

\section{Pricing interest rate derivatives }
In Section~\ref{numericalres}, we study the impact of noisy data in the PCA methodology. In Section~\ref{empan}, we report the existence of nontrivial measurement errors in forward rate markets. In this section, the primary goal is to illustrate the importance of a correct spectral analysis in noisy forward rate curves. The example we choose to illustrate this point is the pricing of interest rate derivatives in the presence of noisy data. We study the impact of observational errors in pricing interest rate derivatives by means of the PCA methodology. More precisely, we compare the pricing error of the PCA methodology based on $\hat{V}_s$ against the LRCM estimators.

Similar to \citet{key-1}, we consider a vanilla-type call option based on a zero coupon bond with 10 years maturity, with time to maturities of .25, .5 and 1 years, and distinct strike prices. The underlying data generating process is specified by a 2-factor Hull-White (g2++) model (see e.g~pages 356-364 in \cite{Soto2007}) which yields an analytic pricing formula for vanilla options written on a zero coupon bond. This model is affine whose the correspondent forward rate volatility function is given by

$$
Vol[df(t,T)]=\sqrt{\upsilon_{1}^{2}e^{-\kappa_{1}(T-t)}+\upsilon_{2}^{2}e^{-\kappa_{2}(T-t)}+2\upsilon_{1}\upsilon_{2}\rho_{12} e^{-(\kappa_{1}+\kappa_{2})(T-t)}};~0\le t\le T,
$$
where the parameters $v_1,v_2,\kappa_1,\kappa_2,\rho_{12} \in \mathbb{R}$. The payoff of interest is given by
$$V_{T^0}: = \max\big\{P(T^0,T) - K,0\big\}$$
for some $T^0 < T$, where $T$ is the time of exercising the payoff $V_{T^0}$. The Gaussian structure of the 2-factor Hull-White model yields a closed form expression for the price $\mathbb{C}^0(T^0,T)$ of $V_{T^0}$ as follows

$$\mathbb{C}^0(T^0,T) = P(0,T)\Phi(d_+) - P(0,T^0)\Phi(d_-),$$
where $\Phi$ is the Gaussian cumulative distribution function, $d_{\pm}  := \frac{ln\frac{P(0,T)}{KP(0,T^0)} \pm \frac{v}{2}}{\sqrt{v}}$ and $v$ is the integrated volatility between $[T^0,T]$ which follows equation~(4.41) in~\citet{Andersen2010}, page (185).

%The integrated volatility $v$ is estimated by means of PCA based on the estimators $\hat{V}_s$, $\hat{V}^A_{lr}$, $\hat{V}^{VK}_{lr}$ and $\hat{V}^{UA(p)}_{lr}$ described in Section~\ref{methods}.

We construct the interest rate curves with maturities 3, 6, 9, 12, 15, 18, 21, 24, 30, 36, 40, 60, 72, 90, 108 and 120 months with delta of order 1/252 (daily data) between each curve observation. In order to verify the robustness of our results, we perform a Monte Carlo study with two distinct set of parameters  - the first one consist $\kappa_{1}=.8, \kappa_{2}=.7, v_{1}=.1, v_{2}=.1$ and $\rho_{12}=-.3$, and the second one $\kappa_{1}=.9, \kappa_{2}=.85, v_{1}=.1, v_{2}=.2$ and $\rho_{12} = -.3$. The Monte Carlo experiments are based on the observational process $Z = y + \eta$, where $y$ is the correspondent yield curve process specified by the above 2-factor Hull-White model and $\eta$ is zero or an IID zero mean Gaussian noise with variance $.0035$. The error is understood in terms of Mean Square Error (MSE)
$$MSE~(Price_{pca(Z)}(T^0,T), \mathbb{C}^0(T^0,T))$$
where $Price_{pca(Z)}(T^0,T)$ is the price computed by means of the estimation of the integrated volatility $v$ based on PCA as a function of the sample covariance matrix $\hat{V}_s$ and the LRCM $\hat{V}^{A}_{lr},\hat{V}^{VK}_{lr}$ and $\hat{V}^{UA(p)_{lr}}$, given, respectively, by formulas~(\ref{maes}), (\ref{vk}) and (\ref{ua}).

Table~\ref{zerotab250} reports the Mean Square Error pricing with the first set of parameters, $\eta=0$ and based on a sample size of 250 observations. The PCA pricing using the sample covariance matrix $\hat{V}_{s}$  and the estimators $\hat{V}_{lr}^{A}$ and $\hat{V}_{lr}^{UA(p)}$ presents similar results in terms of pricing errors, while the $\hat{V}_{lr}^{VK}$ estimator achieves the best results. This can be explained by the excellent finite sample properties of $\hat{V}_{lr}^{VK}$ compared with the other estimators. See~\cite{key-2,key-3} for further details.

The results reported in Tables \ref{tab250} and \ref{alttab250} suggest that the pricing of interest rate derivatives based on PCA with sample covariance matrix $\hat{V}_{s}$ is sensitive to observational errors. Clearly, the pricing based on the LRCM estimators $\hat{V}_{lr}^{A}$, $\hat{V}_{lr}^{VK}$ and $\hat{V}_{lr}^{UA(p)}$ presents better performance. We clearly observe smaller Mean Squared Errors for all strikes and time to maturities. This restates the validity and relevance of the corrections w.r.t the estimation of the covariance matrix in the PCA methodology.

The estimator $\hat{V}_{lr}^{VK}$ presents the best performance among the methods discussed in Section~\ref{methods}. This result is particularly important because it is the simplest one to implement where no bandwidth parameter estimation is needed. These results strongly suggest that neglecting the underlying dependence structure generated by observational errors may degenerate significantly the pricing of interest rate derivatives. We also present in Table \ref{tab1,000} the experiment with the first parameter set and with contaminations errors, but now using a sample of 1,000 observations. The results are similar to Tables  \ref{tab250} and \ref{alttab250}. The Monte Carlo experiment of this section is robust w.r.t sample size and parameters.

\begin{table}[ht]
\begin{center}
\begin{tabular}{rrrrrr}
  \hline
.25 Years to Mat.&  Strike  & 0.45 & 0.5 & 0.55 & 0.6 \\
  \hline
&  Static $\hat{V}_{s}$ & 0.034 & 0.047 & 0.053 & 0.051 \\
&  Andrews $\hat{V}_{lr}^{A}$& 0.033 & 0.046 & 0.053 & 0.050 \\
&  V-K $\hat{V}_{lr}^{VK}$ & 0.011 & 0.017 & 0.020 & 0.019 \\
&  M\"uller  $\hat{V}_{lr}^{UA(p)}$ & 0.033 & 0.046 & 0.052 & 0.050 \\
   \hline
.5 Years to Mat.&  Strike  & 0.48 & 0.53 & 0.58 & 0.63 \\
  \hline
&  Static $\hat{V}_{s}$& 0.012 & 0.022 & 0.031 & 0.033 \\
&  Andrews $\hat{V}_{lr}^{A}$& 0.012 & 0.022 & 0.031 & 0.033 \\
&  V-K $\hat{V}_{lr}^{VK}$& 0.004 & 0.007 & 0.010 & 0.011 \\
&  M\"uller  $\hat{V}_{lr}^{UA(p)}$& 0.012 & 0.022 & 0.031 & 0.032 \\
  \hline
1 Year to Mat.&  Strike  & 0.5 & 0.55 & 0.6 & 0.65 \\
  \hline
&  Static $\hat{V}_{s}$ & 0.005 & 0.013 & 0.021 & 0.024 \\
&  Andrews $\hat{V}_{lr}^{A}$& 0.005 & 0.013 & 0.021 & 0.024 \\
&  V-K $\hat{V}_{lr}^{VK}$& 0.002 & 0.004 & 0.007 & 0.008 \\
&  M\"uller  $\hat{V}_{lr}^{UA(p)}$& 0.006 & 0.013 & 0.021 & 0.024 \\
   \hline
\end{tabular}
\caption{MSE - Option Pricing on a Zero Coupon Bond -  without measurement errors, first parameter set, sample size 250. \label{zerotab250}}
\end{center}
\end{table}

% latex table generated in R 2.15.1 by xtable 1.7-0 package
% Mon Aug 27 10:35:50 2012
\begin{table}[ht]
\begin{center}
\begin{tabular}{rrrrrr}
  \hline
.25 Years to Mat.&  Strike  & 0.45 & 0.5 & 0.55 & 0.6 \\
  \hline
&  Static $\hat{V}_{s}$& 0.056 & 0.069 & 0.076 & 0.075 \\
&  Andrews $\hat{V}_{lr}^{A}$& 0.039 & 0.052 & 0.058 & 0.057 \\
&  V-K $\hat{V}_{lr}^{VK}$& 0.013 & 0.019 & 0.022 & 0.021 \\
&  M\"uller  $\hat{V}_{lr}^{UA(p)}$& 0.034 & 0.046 & 0.051 & 0.049 \\
   \hline
.5 Years to Mat.&  Strike  & 0.48 & 0.53 & 0.58 & 0.63 \\
  \hline
&  Static $\hat{V}_{s}$& 0.037 & 0.049 & 0.059 & 0.062 \\
&  Andrews $\hat{V}_{lr}^{A}$& 0.020 & 0.030 & 0.039 & 0.041 \\
&  V-K $\hat{V}_{lr}^{VK}$& 0.004 & 0.008 & 0.011 & 0.012 \\
&  M\"uller  $\hat{V}_{lr}^{UA(p)}$& 0.013 & 0.022 & 0.030 & 0.032 \\
  \hline
1 Year to Mat.&  Strike  & 0.5 & 0.55 & 0.6 & 0.65 \\
  \hline
&  Static $\hat{V}_{s}$& 0.029 & 0.041 & 0.051 & 0.056 \\
&  Andrews $\hat{V}_{lr}^{A}$& 0.013 & 0.021 & 0.030 & 0.034 \\
&  V-K $\hat{V}_{lr}^{VK}$ & 0.002 & 0.004 & 0.007 & 0.009 \\
&  M\"uller  $\hat{V}_{lr}^{UA(p)}$ & 0.006 & 0.013 & 0.021 & 0.024 \\
   \hline
\end{tabular}
\caption{MSE - Option Pricing on a Zero Coupon Bond - with measurement errors, first parameter set, sample size 250\label{tab250}}
\end{center}
\end{table}

% latex table generated in R 2.15.1 by xtable 1.7-0 package
% Mon Aug 27 10:35:50 2012
\begin{table}[ht]
\begin{center}
\begin{tabular}{rrrrrr}
  \hline
.25 Years to Mat.&  Strike  & 0.45 & 0.5 & 0.55 & 0.6 \\
  \hline
&  Static $\hat{V}_{s}$&0.078 & 0.087 & 0.091 & 0.088 \\
&  Andrews $\hat{V}_{lr}^{A}$&0.066 & 0.076 & 0.078 & 0.075 \\
&  V-K   $\hat{V}_{lr}^{VK}$ & 0.024 & 0.029 & 0.030 & 0.029 \\
&  M\"uller $\hat{V}_{lr}^{UA(p)}$ & 0.060 & 0.069 & 0.071 & 0.068 \\
   \hline
.5 Years to Mat.&  Strike  & 0.48 & 0.53 & 0.58 & 0.63 \\
  \hline
&  Static $\hat{V}_{s}$& 0.048 & 0.058 & 0.065 & 0.065 \\
&  Andrews$\hat{V}_{lr}^{A}$& 0.035 & 0.044 & 0.050 & 0.051 \\
&  V-K   $\hat{V}_{lr}^{VK}$& 0.009 & 0.013 & 0.016 & 0.016 \\
&  M\"uller  $\hat{V}_{lr}^{UA(p)}$& 0.027 & 0.036 & 0.041 & 0.041 \\
  \hline
1 Year to Mat.&  Strike  & 0.5 & 0.55 & 0.6 & 0.65 \\
  \hline
 & Static $\hat{V}_{s}$& 0.036 & 0.046 & 0.054 & 0.056 \\
 & Andrews  $\hat{V}_{lr}^{A}$& 0.022 & 0.031 & 0.037 & 0.039 \\
 &  V-K $\hat{V}_{lr}^{VK}$ & 0.005 & 0.008 & 0.011 & 0.012 \\
 & M\"uller  $\hat{V}_{lr}^{UA(p)}$& 0.015 & 0.022 & 0.027 & 0.029 \\
   \hline
\end{tabular}
\caption{MSE - Option Pricing on a Zero Coupon Bond - with measurement errors, second parameter set,  sample size 250.  \label{alttab250}}
\end{center}
\end{table}

% latex table generated in R 2.15.1 by xtable 1.7-0 package
% Mon Aug 27 10:35:50 2012
\begin{table}[ht]
\begin{center}
\begin{tabular}{rrrrrr}
  \hline
.25 Years to Mat.&  Strike  & 0.45 & 0.5 & 0.55 & 0.6 \\
  \hline
&  Static $\hat{V}_{s}$& 0.059 & 0.071 & 0.076 & 0.073 \\
&  Andrews $\hat{V}_{lr}^{A}$& 0.039 & 0.050 & 0.054 & 0.051 \\
&  V-K  $\hat{V}_{lr}^{VK}$& 0.009 & 0.014 & 0.016 & 0.016 \\
&  M\"uller $\hat{V}_{lr}^{UA(p)}$ & 0.033 & 0.044 & 0.047 & 0.044 \\
\hline
.5 Years to Mat.&  Strike  & 0.48 & 0.53 & 0.58 & 0.63 \\
  \hline
&  Static $\hat{V}_{s}$& 0.039 & 0.051 & 0.060 & 0.062 \\
&  Andrews $\hat{V}_{lr}^{A}$& 0.018 & 0.028 & 0.035 & 0.036 \\
&  V-K $\hat{V}_{lr}^{VK}$ & 0.002 & 0.005 & 0.008 & 0.009 \\
&  M\"uller $\hat{V}_{lr}^{UA(p)}$ & 0.013 & 0.022 & 0.028 & 0.029 \\
  \hline
1 Year to Mat.&  Strike  & 0.5 & 0.55 & 0.6 & 0.65 \\
  \hline
&  Static $\hat{V}_{s}$& 0.031 & 0.043 & 0.052 & 0.056 \\
&  Andrews $\hat{V}_{lr}^{A}$& 0.011 & 0.019 & 0.027 & 0.029 \\
&  V-K $\hat{V}_{lr}^{VK}$& 0.001 & 0.003 & 0.005 & 0.006 \\
&  M\"uller $\hat{V}_{lr}^{UA(p)}$& 0.006 & 0.013 & 0.019 & 0.021 \\
\hline
\end{tabular}
\caption{MSE - Option Pricing on a Zero Coupon Bond - with measurement error,  first parameter set, sample size 1,000\label{tab1,000}}
\end{center}
\end{table}

\section{Final remarks}\label{conc}

In this article, we discuss the impact of the inherent presence of measurement errors in the classical application of the PCA methodology in the estimation of forward rate curves. Our results strongly suggest the classical PCA method based on the standard sample covariance matrix is \textit{not} suitable for estimating factors of forward rate curves. The main reason is the appearance of non-negligible observational errors in the implied forward rate curves. An alternative methodology based on so-called long-run covariance matrix estimators seems to improve significantly the quality of the estimation of the principal components of forward rate curves. We illustrate the importance of a correct spectral analysis in forward rate curves by presenting non-trivial effects of noisy data in pricing errors related to European call options. Lastly, the results of this paper yield a sound explanation for the remarkable difference between the estimated number of factors in yield and forward rate curves reported in the literature (see e.g.~\cite{akahori},~\cite{akahori1},~\cite{lekkos} and~\cite{liu}).

Our conclusion is supported by the following results presented in this paper. Proposition~\ref{numberres} proves that the number of principal components of the observed forward rate and yield curves must be identical in the absence of measurement errors. Based on this fact, we perform a detailed simulation analysis in three prominent interest rate models with a number of distinct parameters in the presence of observational errors of various magnitudes and forms. Market microstructure effects and interpolation errors due to extraction of forward rates from yield curves are carefully analyzed. The results clearly indicate a considerable bias in the classical PCA methodology applied to forward rate curves. In contrast, Monte Carlo experiments jointly with empirical analysis strongly suggest that PCA based on long run covariance matrices is robust w.r.t measurement errors in forward rate curves. The presence of observational errors in forward rate markets is validated by Proposition~\ref{numberres} together with a detailed empirical analysis on the number of principal components in the UK and US interest rate markets. Monte Carlo experiments reports that classical PCA based on sample covariance matrices presents nontrivial pricing errors for European call options. In contrast, the use of long run covariance matrices seems to correct this bias.

 %Lastly, we illustrate the importance of the observational errors in pricing interest rate derivatives. In particular, we show that PCA based on long range covariance matrices yields a significant reduction in the pricing errors due to noisy data.

\bibliographystyle{elsarticle-harv}
\bibliography{bibejor}

\end{document}